\documentclass[preprint]{raa} 

\usepackage{graphicx,times}             
\usepackage{natbib}
\usepackage{tabularx}
\usepackage{amssymb,amsmath}
\usepackage{siunitx}
\usepackage[pagebackref=true]{hyperref}

\newcommand{\orcid}[1]{
    \raisebox{0.7ex}{\scalebox{1}{
        \href{https://orcid.org/#1}{\includegraphics[height=1.5ex]{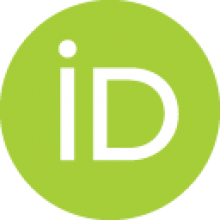}}
    }}
}

\begin{document}

  \title{\textit{SVOM}/VT: Preliminary Calibration Analysis}

   \volnopage{Vol.0 (20xx) No.0, 000--000}      
   \setcounter{page}{1}          

   \author{Zhu-Heng Yao\orcid{0009-0000-1228-2373}\inst{1}\thanks{Corresponding author. Email: zhyao@bao.ac.cn} 
   \and Yu-Lei Qiu\orcid{0009-0007-7207-4884}\inst{1}\thanks{Corresponding author. Email: qiuyl@nao.cas.cn}
   \and Jin-Song Deng\orcid{0000-0001-5646-8583}\inst{1,2}\thanks{Corresponding author. Email: jsdeng@nao.cas.cn}
   \and Li-Ping Xin\orcid{0000-0002-9422-3437}\inst{1}
   \and Chao Wu\orcid{0009-0001-7024-3863}\inst{1}
   \and Hua-Li Li\orcid{0000-0002-7457-4192}\inst{1}
   \and Jing Wang\inst{1}
   \and Yi-Nuo Ma\orcid{0009-0007-1390-7575}\inst{1,3,4}
   \and Hong-Bo Cai\inst{1}
   \and Xu-Hui Han\orcid{0000-0002-6107-0147}\inst{1}
   \and Jian-Yan Wei\inst{1,2}
   \and Betrand Cordier\inst{5}
   }

   \institute{National Astronomical Observatories, Chinese Academy of Sciences,
             Beijing 100012, China; {\it zhyao@bao.ac.cn, qiuyl@nao.cas.cn,
             jsdeng@nao.ac.cn}\\
        \and
             School of Astronomy and Space Science, University of Chinese Academy of Sciences, Beijing 101408, China\\
        \and 
             Institute for Frontier in Astronomy and Astrophysics, Beijing Normal University, Beijing 102206, People's Republic of China
        \and
             School of Physics and Astronomy, Beijing Normal University, Beijing 100875, People's Republic of China
        \and
             CEA/Paris-Saclay, Irfu/D\'epartement d'Astrophysique, 91191 Gif-sur-Yvette, France \\
\vs\no
   {\small Received 20xx month day; accepted 20xx month day}}

\abstract{
We present the in-orbit calibration of the Visible Telescope (VT), one of the key instruments aboard the Space Variable Objects Monitor (\textit{SVOM}) mission for gamma-ray burst (GRB) studies.
Using \textit{Gaia}~Data Release 3 (DR3) as a reference, the VT achieves an astrometric precision better than $0.03''$ for bright stars, degrading to $\sim 0.25''$ for faint targets.
Shortly after launch, contamination was detected, reducing system transmission by $\sim40\%$.
An initial bake-out successfully restored performance, but gradual recontamination caused transmission to decline by $\sim20\%$ over the following 100 days before stabilizing. Despite this effect, routine standard star observations maintain precise zero-point calibration, ensuring a photometric stability of $0.02$ mag.
Using synthetic stellar spectra, we derived photometric transformations to the \textit{Gaia}, SDSS, and Johnson-Cousins systems with typical residuals of $0.03$ mag.
These results demonstrate the VT system's capability and reliability in calibrating GRBs and other transients.
}

   \authorrunning{Z.-H. Yao, Y.-L. Qiu \& J.-S. Deng, et al.}
   \titlerunning{VT Calibration}  

   \maketitle

%
%
\section{Introduction}
\label{sec:intro}

The \textit{SVOM} is a Chinese-French joint mission dedicated to the study of GRBs and other transient phenomena in the high-energy universe~\citep{Wei2016,Cordier2026}.
The VT onboard \textit{SVOM} is a 44-cm Ritchey--Chr\'etien optical telescope designed to provide rapid and sensitive follow-up observations of GRBs detected and localized by the \textit{SVOM}/ECLAIRs~\citep{Godet2026}.
Using a dichroic beam splitter, VT simultaneously observes in the blue band ($0.40-0.65\,\mu$m; hereafter VT$\_B$) and the red band ($0.65-1.00\,\mu$m; hereafter VT$\_R$), enabling prompt two-color photometry over a $26'\times26'$ field of view (FOV).
Each band is equipped with a $2\mathrm{k}\times2\mathrm{k}$ e2V CCD detector and is required to reach a limiting magnitude of $\sim22.5$ in a 300~s exposure~\citep{Qiu2026}.

A central objective of VT within the \textit{SVOM} mission is to refine GRB localization to sub-arcsecond levels in celestial coordinates to facilitate timely redshift measurements by ground-based large telescopes, necessitating precise astrometric calibration.
In addition, VT aims to construct a homogeneous and well-sampled dataset of optical light curves of GRB afterglows, capturing detailed temporal features such as early brightenings, plateaus, flares, late breaks and rebrightenings.
These temporal features provide key insights into the nature of GRB central engines, radiation mechanisms, and shock physics~\citep{Zhang2007_GRBswift}.
Furthermore, the color information derived from the two VT bands, particularly deep upper limits in instances of non-detection, may serve as a preliminary indicator of high-redshift GRBs~\citep{Wang2020,Qiu2026}.
All of these scientific objectives critically rely on accurate and stable astrometric and photometric calibration.

To mitigate atmospheric turbulence and gravitational flexure, current ground-based wide-field surveys such as DES~\citep{Abbott2018_DESDR1}, Pan-STARRS~\citep{Magnier2020_PScali}, and LSST~\citep{Ivezic2019_LSST} rely on \textit{Gaia}-based reference catalogs~\citep{Gaia2016} to establish astrometric solutions at the level of tens of milli-arcseconds (mas).
Furthermore, nightly observations of standard stars are performed to maintain photometric zero points with accuracies of a few percent.

For a space-borne optical telescope like the VT, although atmospheric turbulence and terrestrial gravitational flexure are absent, regular onboard calibrations are necessary to compensate for the low-frequency evolution of thermoelastic deformations, ongoing degradation of optical coatings and detectors due to radiation damage, and accumulated outgassing contamination on sensitive optical surfaces~\citep{Poole2008_SwiftUVOTphotcali,Massey2010_HST_CTI,Gardner2023_JWST,Euclid2023_IceContamination2023}.
For example, the Hubble Space Telescope (\emph{HST}; \citealt{Holtzman1995,Sirianni2005}) utilizes a mature geometric distortion model calibrated through repeated observations of dense stellar fields, achieving sub-10 mas astrometric precision.
The \emph{Swift}/UVOT instrument~\citep{Poole2008_SwiftUVOTphotcali} applies a stable distortion map and uses UV spectrophotometric standard stars to maintain its long-term photometric accuracy.
More recently, the \emph{Euclid} mission~\citep{Euclid2025_overview} anchors its global astrometric solution directly to the \textit{Gaia}~DR3 catalog and employs a large set of high-quality standard stars for absolute flux calibration, achieving mas-level internal consistency.

The VT$\_B$ and VT$\_R$ bands are calibrated to the AB magnitude system.
They have similar bandwidths and crossover wavelengths to those of \textit{Gaia}'s $BP$ and $RP$ bands, but
differ significantly from either the Sloan or Johnson-Cousins photometric systems widely used in ground-based GRB follow-up observations.
To enable coherent multi-instrument analyses of optical light curves from VT and other facilities, establishing reliable photometric transformations between the VT system and other standard systems (in particular, Sloan and Johnson-Cousins) is critical.
Moreover, a precise determination of the effective wavelength for each band is essential for constructing broadband GRB spectral energy distributions (SEDs).

The structure of this paper is as follows.
Section~\ref{sec:astro_cali} describes the astrometric calibration procedure and the accuracy achieved.
Section~\ref{sec:photo_cali} presents the photometric calibration framework, detailing system response curves (Sect.~\ref{subsec:response_curve}), photometric zero-points (Sect.~\ref{subsec:zp}), long-term stability monitoring using standard spectrophotometric stars (Sect.~\ref{subsec:stds}), and photometric system transformations (Sect.~\ref{subsec:trans}).
Section~\ref{sec:dis} discusses potential contamination effects and the red and blue leakage in both bands.
Finally, Section~\ref{sec:sum} summarizes the main results of this work.

\section{Astrometric Calibration}
\label{sec:astro_cali}

\subsection{Astrometric Calibration Procedure}
\label{subsec:astro_cali_proc}
The astrometric calibration establishes the transformation between detector pixel coordinates and the International Celestial Reference System (ICRS).
An initial pointing solution is obtained from the onboard spacecraft attitude information and subsequently refined using reference stars from \textit{Gaia}~DR3 to achieve high-precision astrometry.

The calibrated spacecraft attitude is 
as a quaternion describing the orientation of the VT optical axis, which is aligned with the FOV center of the red  channel.
This information yields the nominal pointing coordinates ($RA_0$, $Dec_0$) and the roll angle $\phi_0$.
Combined with the detector pixel scale, these parameters define an initial approximate World Coordinate System (WCS) for each VT image, which enables the projection of proper-motion-corrected \textit{Gaia}~DR3 reference stars onto the detector plane.
The accuracy of this initial WCS is constrained by that of the attitude calibration (performed between the VT and star trackers), which is $\sim15''$ at $90\%$ confidence level~\citep{Qiu2026}.

The initial WCS solution is refined by matching bright VT sources with well-determined centroids to the projected \textit{Gaia}~DR3 positions using a search radius of 20 pixels.
An intermediate astrometric solution is derived with the {\sc iraf}~\citep{Tody1986,Tody1993} task \texttt{ccmap}, using typically 20--30 anchor stars (see Fig.~2 of \citealt{Wu2026} for the geometric schematic) selected as the brightest unsaturated sources in each exposure.
A second- or third-order polynomial model is adopted to correct for large-scale field-dependent distortions, and the fitting is performed iteratively with outlier rejection until a stable sub-arcsecond solution is achieved.
Based on this intermediate solution, the fitting procedure is repeated using an expanded sample of more than 100 bright sources and a tighter matching tolerance at the sub-arcsecond level.
A fourth-order or higher polynomial model is then applied to achieve  better residual systematics and small-scale geometric distortions.
The resulting optimized solution achieves an astrometric accuracy of approximately $0.01''$.

In rare cases (e.g., crowded fields) where the initial attitude-based WCS fails to yield good alignment with \textit{Gaia} projections, a blind astrometric solution is derived using {\sc astrometry.net}~\citep{Lang2010}.
However, {\sc astrometry.net} is employed only as a backup option in our procedure by taking into accounts of computational costs.

\subsection{In-orbit Performance of Astrometric Calibration}
\label{subsec:astro_perform}

\begin{figure}
\centering
\includegraphics[width=0.45\textwidth, angle=0]{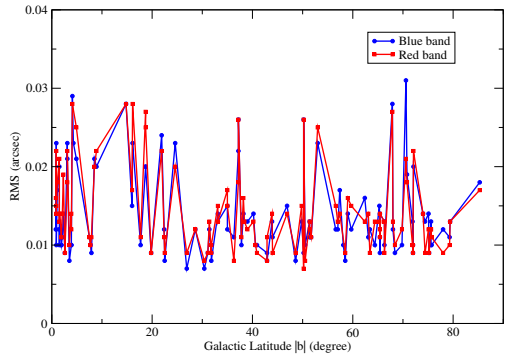}
\caption{VT astrometric calibration residuals (RMS) versus Galactic latitude in the 110 GRB fields.
}
\label{fig:rms_gal_latitude}
\end{figure}

The in-orbit astrometric performance of VT is evaluated using 110 GRB fields observed during routine operations.
As shown in Figure~\ref{fig:rms_gal_latitude}, more than $90\%$ of the analyzed fields achieve an astrometric precision better than $0.03''$, fully satisfying the requirements for the rapid detections of GRB counterparts and follow-up observations.
The root-mean-square (RMS) of astrometric residuals shows no significant correlation with Galactic latitude, despite localized variations ($\sim0.01''-0.03''$).
Both bands exhibit consistent systematic trends, with only minor discrepancies.
Considering the two bands operate independently, their synchronous variations cannot be attributed to noise, but shall be from a common systematic source: either stellar density or pointing stability.
However, the absence of a correlation between Galactic latitude (a proxy for stellar density) and astrometric accuracy excludes stellar density as the cause.
We therefore conclude that variations in astrometric accuracy are most likely driven by spacecraft pointing stability.

Furthermore, to extend our assessment of VT astrometric performance to fainter sources, we analyze the dependence of positional residuals on stellar brightness using a representative VT image.
We additionally employ the Legacy Survey DR10 (LS~DR10;~\citealt{Dey2019_LSoverview}) catalog as an independent reference, which offers substantially deeper magnitude coverage.

\begin{figure}
\centering
\includegraphics[width=0.45\textwidth, angle=0]{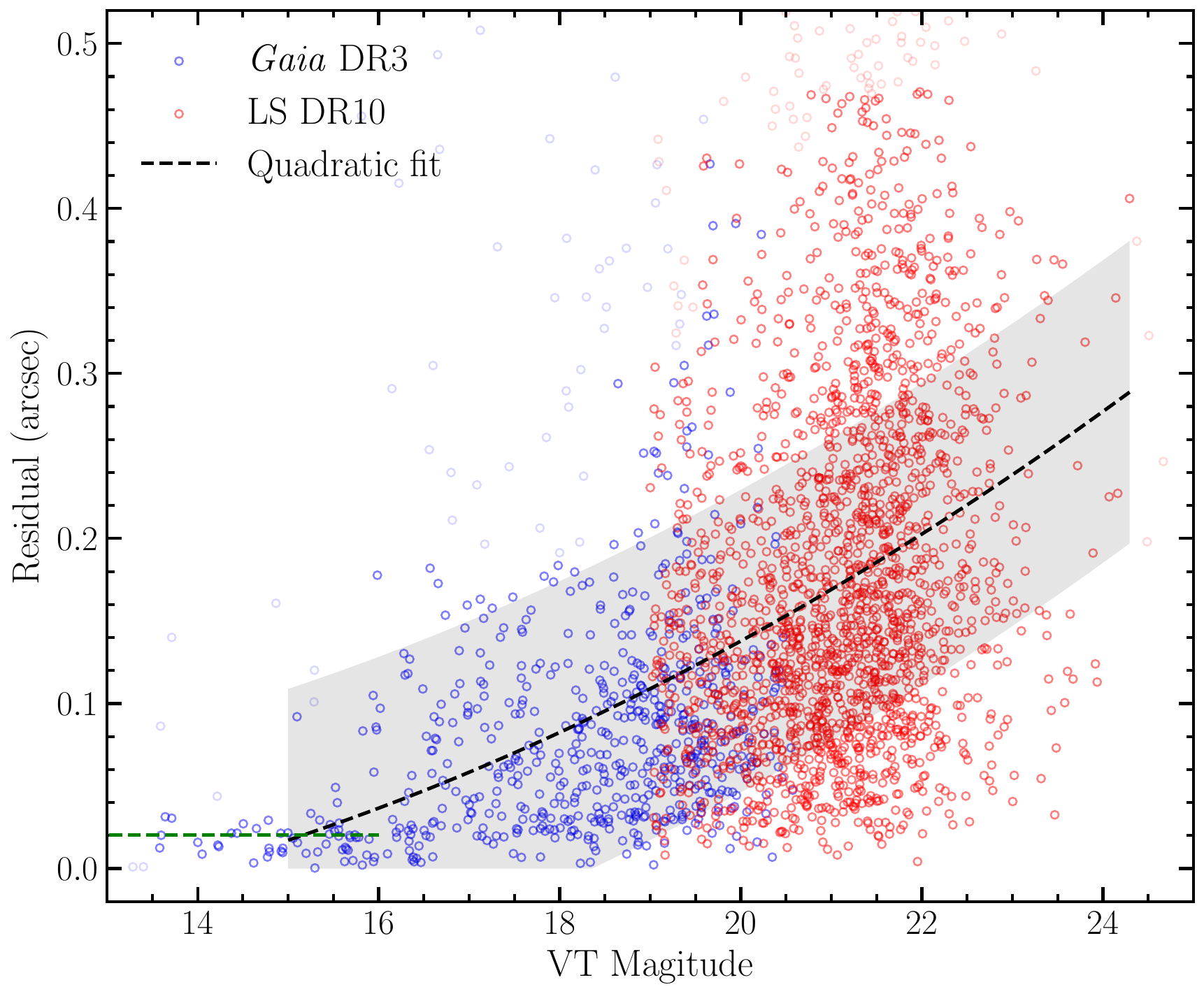}
\caption{Relationship between stellar magnitude and astrometric residuals (blue circles for \textit{Gaia}~DR3 stars and red circles for LS~DR10 stars).
The green horizontal dashed line indicates the astrometric residual $0.02''$ measured for bright stars.
The black dashed curve represents the best-fitting second-order polynomial to the data, and the gray shaded region indicates the corresponding 1-$\sigma$ uncertainty range.
}
\label{fig:res_vs_mag}
\end{figure}

The results are illustrated in Figure~\ref{fig:res_vs_mag}.
For bright stars, the astrometric residuals (RMS) is $\sim0.02''$.
The astrometric residuals increase gradually toward fainter magnitudes.
To quantitatively describe this trend, we fit the residual–magnitude relationship with a second-order polynomial function.
Within the magnitude range effectively sampled by LS~DR10, down to approximately 23 mag, the positional accuracy (RMS) of VT remains with a mean value of $0.25''$.
These results demonstrate that VT maintains reliable astrometric performance even for faint sources, ensuring accurate localization of GRB optical counterparts and other faint transients.

\section{Photometric Calibration}
\label{sec:photo_cali}

\subsection{Spectral Characteristics of VT Bands}
\label{subsec:response_curve}
\begin{figure}
\centering
\includegraphics[width=0.45\textwidth, angle=0]{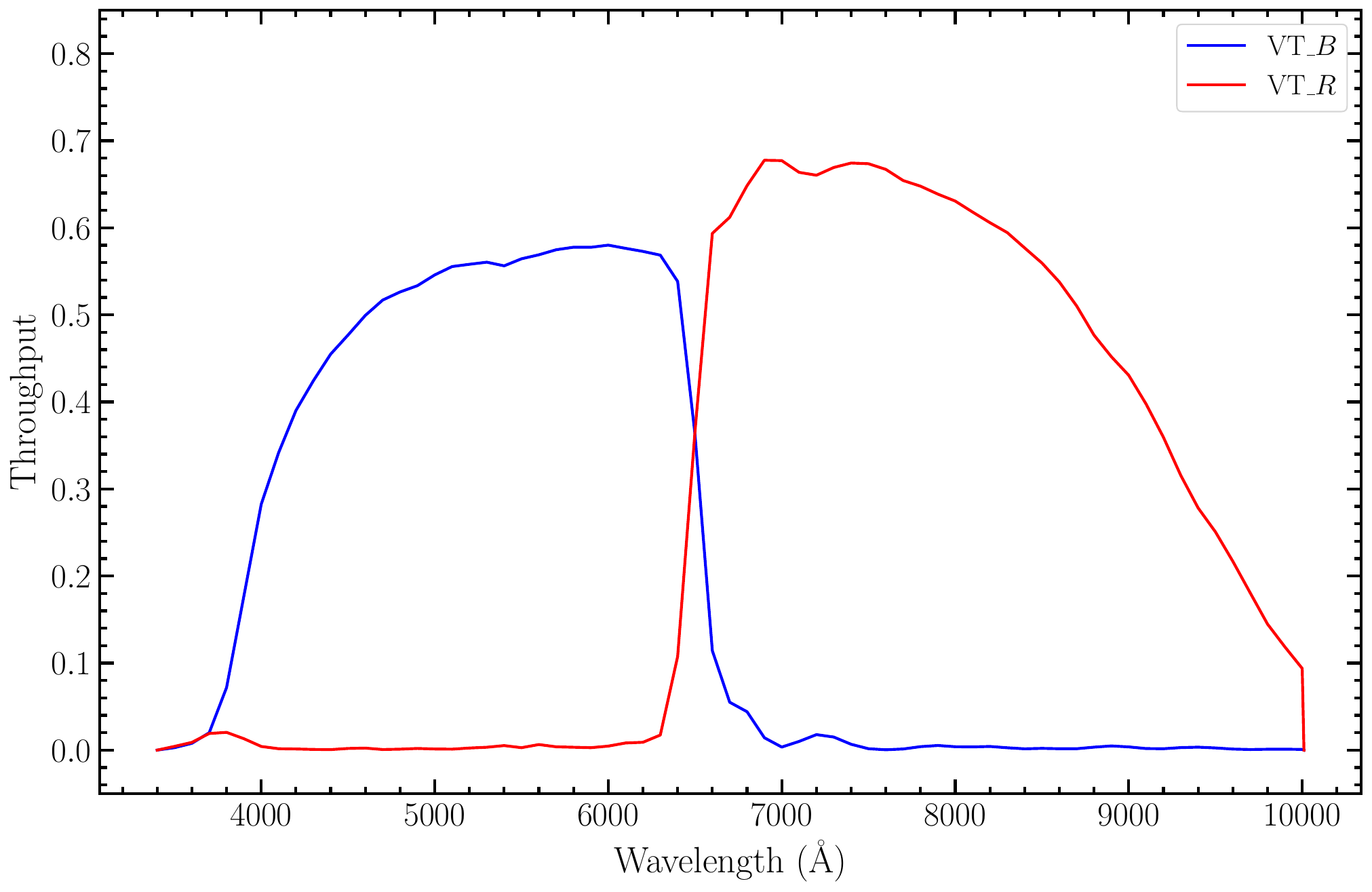}
\caption{The total system response curves of the VT$\_B$ and VT$\_R$ bands.}
\label{fig:vt_filters}
\end{figure}

The wavelength-dependent system response curves, $S(\lambda)$, for the VT$\_B$ and VT$\_R$ bands represent the full optical throughput of the instrument. 
They are constructed by combining the detector quantum efficiency provided by the manufacturer, the reflectivities of the mirrors, the transmission efficiency of the dichroic beam splitter, and other optical efficiency factors (e.g., obscuration and optical losses) along the light path (see~\citealt{Qiu2026,Zhang2026}). 
The resulting response curves are shown in Figure~\ref{fig:vt_filters}.
Key system parameters, defined below from $S(\lambda)$ and summarized in Table~\ref{tab:filter_paras}, are used in the subsequent analyses.

The effective wavelength, $\lambda_{\rm eff}$, defined as the photon-weighted mean wavelength following~\citep{Schneider1983,Fukugita1996_SDSSsystem}, is given by:
\begin{equation}
\label{eq:lambda_eff}
\lambda_{\rm eff} = \exp \left[ \frac{\int \lambda\ln\lambda \,S(\lambda) d\lambda}{\int \lambda S(\lambda)d\lambda} \right].
\end{equation}
The central wavelength defined in this work is consistent with
~\citet{Euclid2022_NISP}:
\begin{equation}
\label{eq:lambda_cen}
    \lambda_{\rm cent}=\frac{\int\lambda S(\lambda)d\lambda}{\int S(\lambda)d\lambda}. 
\end{equation}
The equivalent bandwidth, $\Delta\lambda_{\rm eff}$, corresponds to the width of a rectangular bandpass that transmits the same total flux when multiplied by the throughput at the effective wavelength:
\begin{equation}
    \Delta\lambda_{\rm eff}=\frac{\int S(\lambda) d\lambda}{S(\lambda_{\rm eff})},  
\end{equation}
where $S(\lambda_{\rm eff})$ is the system response at the effective wavelength.
The RMS bandwidth is calculated by:
\begin{equation}
\sigma_{\lambda} = \sqrt{\frac{\int (\lambda - \lambda_{\mathrm{eff}})^2 S(\lambda) d\lambda}{\int S(\lambda) d\lambda}},
\end{equation}
which quantifies the dispersion of the filter transmission profile around its effective wavelength.

As summarized in Table~\ref{tab:filter_paras}, the VT$\_B$ and VT$\_R$ bands have $\lambda_{\rm eff}=539.48$~nm and $794.73$~nm, with central wavelength $\lambda_{\rm cent}=533.67$~nm and $789.09$~nm, respectively.
In both bands, $\lambda_{\rm eff}$ is shifted to slightly longer wavelengths than $\lambda_{\rm cent}$ (by $\sim 5-6$~nm), indicating that the long-wavelength side of the bandpass contributes marginally more to the response-weighted representative wavelength.
The two filters are broad, with ${\rm FWHM}=251.66$~nm (VT$\_B$) and $275.76$~nm (VT$\_R$); the effective bandwidths are $249.91$~nm (VT$\_B$) and $284.96$~nm (VT$\_R$).
The close agreement between $\Delta\lambda_{\rm eff}$ and FWHM for VT$\_B$ suggests a relatively regular single-peaked passband, whereas the slightly larger $\Delta\lambda_{\rm eff}$ than FWHM for VT$\_R$ implies a greater contribution from the low-throughput wings.
The RMS bandwidths are $\sigma_{\lambda}=78.78$~nm for VT$\_B$ and $93.78$~nm for VT$\_R$, indicating that VT$\_R$ has a broader wavelength spread around $\lambda_{\rm eff}$ than VT$\_B$ in the RMS sense.
Finally, the response evaluated at the effective wavelength is close to the peak response, with $S(\lambda_{\rm eff})/S_{\rm peak}\approx0.96$ (VT$\_B$) and $\approx0.94$ (VT$\_R$), showing that $\lambda_{\rm eff}$ lies near the high-throughput region of each passband.

\begin{table}[htbp]
\centering
\caption{Filter Parameters for the VT$\_B$ and VT$\_R$ Bands.}
\label{tab:filter_paras}
\begin{tabularx}{0.45\textwidth}{XXXc}
  \hline\noalign{\smallskip}
Parameter & VT$\_B$ & VT$\_R$ & unit \\
  \hline\noalign{\smallskip}
$\lambda_{\rm eff}$ & 539.48 & 794.73 & nm \\
$\lambda_{\rm cent}$ & 533.67 & 789.09 & nm \\
FWHM & 251.66 & 275.76 & nm \\
$\Delta\lambda_{\rm eff}$ & 249.91 & 284.96 & nm \\
$\sigma_{\lambda}$ & 78.78 & 93.78 & nm \\
$S(\lambda_{\rm eff})$ & 0.557 & 0.635 & - \\
$S_{\rm peak}$ & 0.580 & 0.678 & - \\
  \hline\noalign{\smallskip}
\end{tabularx}
\end{table}

\subsection{Pre-launch Photometric Zero Points}
\label{subsec:zp}

The VT photometric system is defined in the AB magnitude system, following the standard AB magnitude formalism~\citep{Schneider1983,Fukugita1996_SDSSsystem}:
\begin{equation}
    m_{\rm AB} = -2.5 \log_{10} \left[\frac{\int d(\ln\nu) f_{\nu } S({\nu})}{\int d(\ln\nu) S({\nu})}\right] \;-\; 48.60,
\end{equation}
where $f_\nu$ is the flux density in units of $\rm erg\,s^{-1}\,cm^{-2}\,Hz^{-1}$ and $S(\nu)$ denotes the dimensionless system response in the frequency domain.

In practical observations, the AB magnitude corresponds to the measured photon count rate $\rm CR$ as: 
\begin{equation}
    m_{\rm AB} = -2.5 \log_{10}  (\rm CR) + {\rm ZP},
\end{equation}
where $\rm ZP$ is the zero point of the system defined as the corresponding magnitude for 1 photoelectron per second and $\rm CR$ is determined by the incident photon flux and the system response: 
\begin{equation}
    {\rm CR} = A \int F^{\rm phot}_\lambda \, S(\lambda)\, d\lambda,
\end{equation}
where $A=\pi (D/2)^2$, is the collecting area of the telescope, and $F^{\rm phot}_\lambda = f_\lambda / (h\nu)$ is the photon flux density, with $h$ denoting the Planck constant.
Under these definitions, $\rm ZP$ can be expressed as:
\begin{equation}
    {\rm ZP} = 2.5 \log_{10} \left[ A \int S(\lambda)\, d(\ln \lambda) \right] + 16.85.
\end{equation}

Using the system response presented above and adopting a telescope diameter of $D=44$~cm, we derive the AB zero points: 23.37~mag for VT$\_B$ and 23.22~mag for VT$\_R$.
These values are derived from the ground-base calibration.

\begin{table*}[htbp]
\centering
\caption{Parameters for two Standard Stars.}
\label{tab:stds_paras}
\begin{tabularx}{0.9\textwidth}{lXXccc}
\hline\noalign{\smallskip}
Object name &R.A. (J2000.0) & Dec. (J2000.0) & Stellar type & Mag (VT$\_B$) & Mag (VT$\_R$) \\ 
\hline\noalign{\smallskip}
BD+28$^\circ$4211 & 21:51:11.02 & +28:51:50.37 & sdO & 10.37 & 11.19 \\
Feige~34 & 10:39:36.74 & +43:06:09.21 & sdO & 11.02 & 11.78 \\
\hline\noalign{\smallskip}
\end{tabularx}
\end{table*}

\subsection{Zero Point Monitoring in Orbit}
\label{subsec:stds}
\begin{figure}
\centering
\includegraphics[width=0.45\textwidth, angle=0]{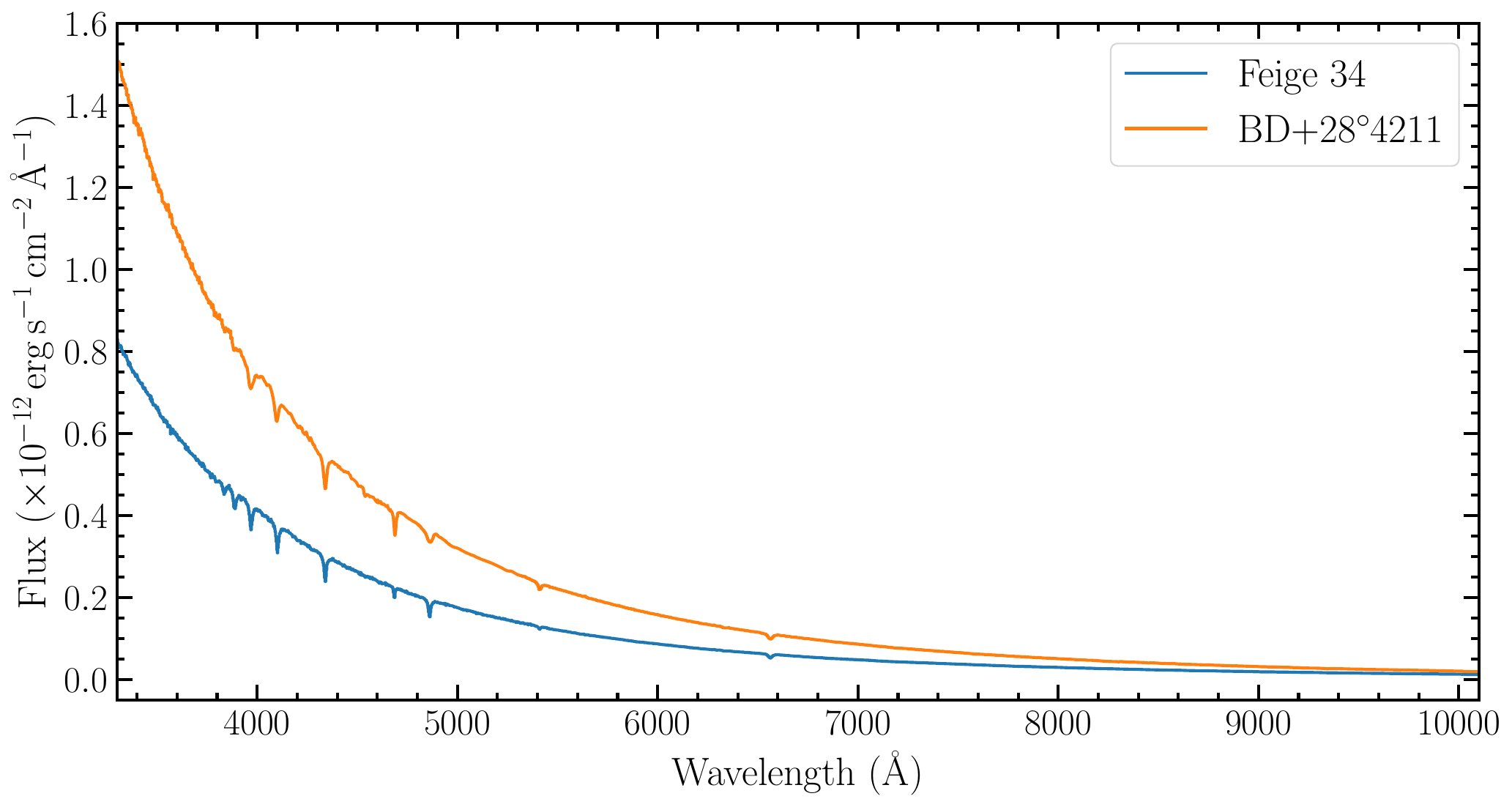}
\caption{Spectra of the two spectrophotometric standard stars BD+28$^\circ$4211 and Feige~34. 
}
\label{fig:stds}
\end{figure}

To track the temporal evolution of the VT zero point, we conducted long-term monitoring of standard stars.
Two bright and stable spectrophotometric standards, BD+28$^\circ$4211 and Feige~34, were selected for this purpose.
Both stars are hot subdwarfs of spectral type sdO~\citep{Massey1990}, with $T_{\rm eff}\approx82{,}000$~K for BD+28$^\circ$4211 \citep{Latour2013_BD28} and $T_{\rm eff}\approx62{,}550$~K for Feige~34 \citep{Latour2018_Feige34}.
Their spectra (see Fig.~\ref{fig:stds}) are drawn from the CALSPEC database\footnote{https://www.stsci.edu/hst/instrumentation/reference-data-for-calibration-and-tools/astronomical-catalogs/calspec}\citep{Bohlin2014,Bohlin2020}, and when combined with the VT system response curves, they provide reference magnitudes for calibration (see Table\ref{tab:stds_paras}).

Standard stars were observed with fixed exposure times and identical CCD readout modes, at a cadence ranging from daily to monthly, ensuring that measurements across different epochs are directly comparable.  
Raw images were first corrected for instrumental effects, including bias subtraction, dark current correction, and flat-fielding, to ensure a uniform detector response.
Aperture photometry was then performed to extract stellar fluxes, adopting full aperture with a fixed radius of 10 pixels ($7.6''$), while the sky background was estimated from an annulus spanning 15--20 pixels ($11.4''-15.2''$) around each source (see~\citealt{Li2026} for VT-specific details).
Given the stable point spread function in VT images~\citep{Qiu2026}, this uniform reduction procedure ensures that the measured flux variations reflect genuine instrumental throughput changes rather than data-processing artifacts.

\begin{figure*}
\centering
\includegraphics[width=0.96\textwidth, angle=0]{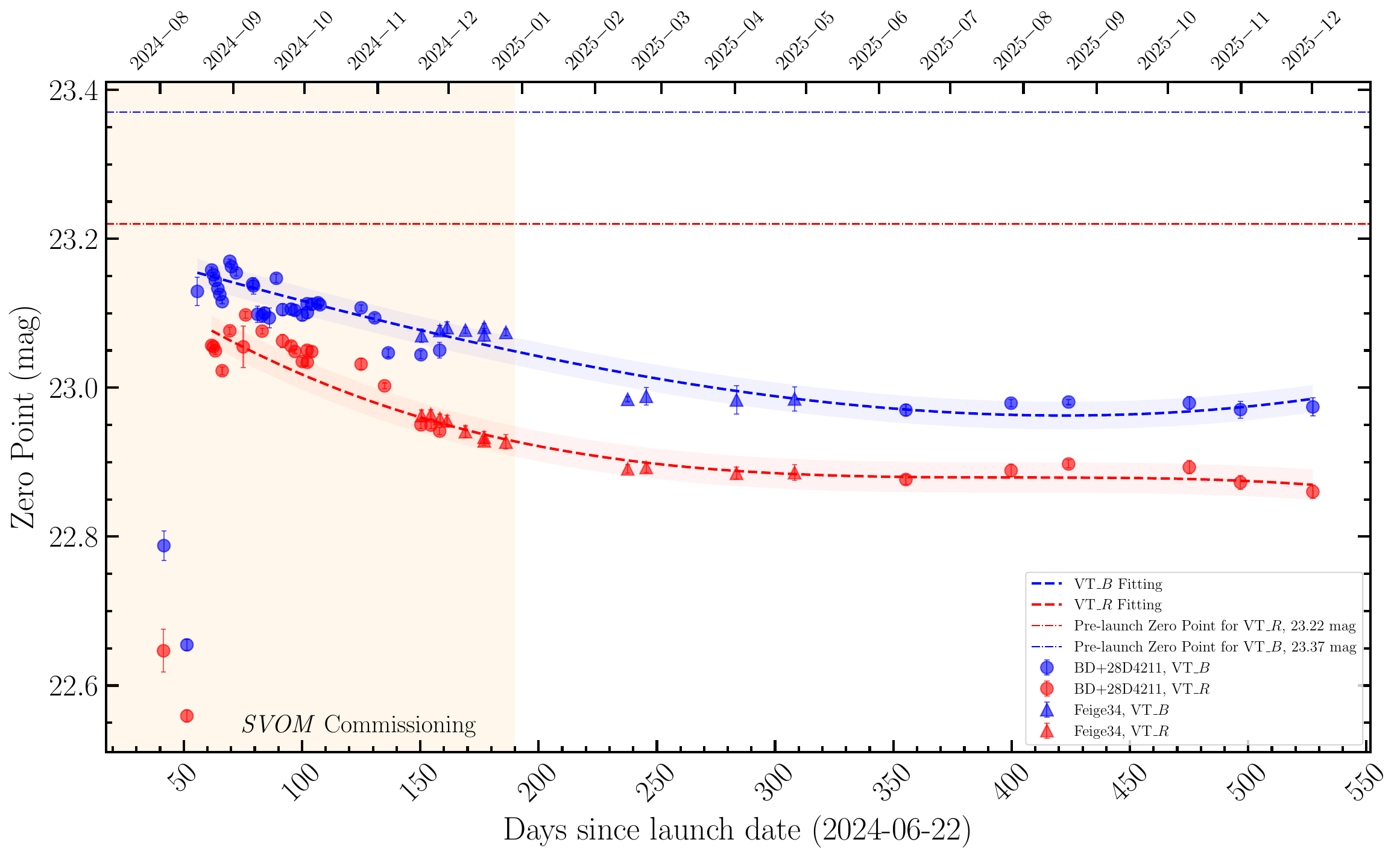}
\caption{Long-term monitoring of the VT$\_R$ (red) and VT$\_B$ (blue) zero points measured from the two standard stars BD+28$^\circ$4211 (circles) and Feige~34 (triangles).
The dash-dot line horizontal lines indicate the ground-based calibration zero points (23.37~mag for VT$\_B$, 23.22~mag for VT$\_R$).
The dashed lines with error bars show a third-order fit to the zero-point evolution.
The orange-shaded region marks the \textit{SVOM} commissioning phase, while the period before it corresponds to the initial pre-launch phase when contamination was most severe.
After commissioning, the zero points remain largely stable, varying by only $\sim2\%$.
}
\label{fig:stds_lc}
\end{figure*}

Figure~\ref{fig:stds_lc} presents the long-term evolution of the VT photometric zero points derived from regular monitoring of two spectrophotometric standard stars.
Shortly after launch, a significant degradation of the photometric zero point was detected in both bands, accompanied by notable halos around the bright objects in the VT images, which is likely related to instrumental contamination.
To mitigate this effect, CCD bake-out operations were immediately performed, which proved highly effective: the first bake-out restored the system transmission by approximately $40\%$, corresponding to a zero-point shift of $\Delta m\sim0.4$.
Despite this recovery, the post–bake-out zero points remain systematically lower than those measured during pre-launch ground calibration by $\sim 10-20\%$.
During the subsequent four months of 2024, the measured zero points exhibited a gradually degrading of approximately 0.2~mag, indicating a renewed contamination.
Since early 2025, the photometric zero points have remained largely stable, showing only minor fluctuations at the $\pm2\%$ level in both bands.
The detailed discussion for the evolution of contamination is presented in Section~\ref{subsec:contamination}.
In the future, the zero point will continue to be tracked through regular observations of spectrophotometric standard stars to maintain the reliability of VT’s scientific measurements.

\subsection{Transformation of the VT Photometric System to External Systems}
\label{subsec:trans}

\begin{figure}
\centering
\includegraphics[width=0.45\textwidth, angle=0]{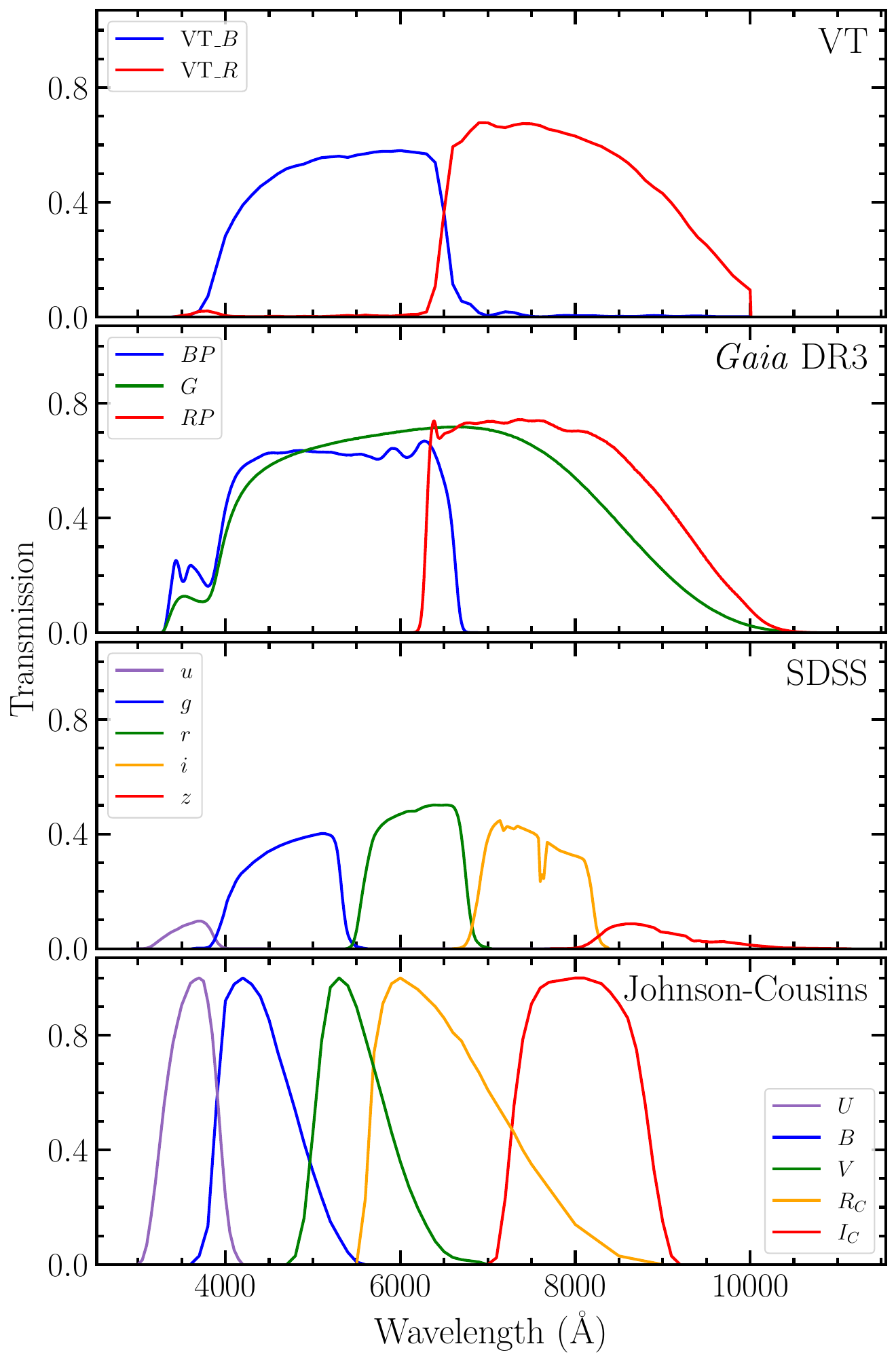}
\caption{Comparison between the VT response curves and those of the \textit{Gaia}, SDSS and Johnson--Cousins systems.}
\label{fig:cmp_filters}
\end{figure}

Reliable transformations between the VT photometric system and widely used external systems are essential for combining VT measurements with existing survey data and for interpreting VT photometry within a broader astrophysical context.
In particular, expressing VT magnitudes in the \textit{Gaia}, SDSS, and Johnson--Cousins systems enables direct comparison with archival catalogs and places VT observations on a well-established photometric scale.

Figure~\ref{fig:cmp_filters} compares the VT passbands with those of \textit{Gaia}, SDSS, and Johnson--Cousins.
The VT$\_B$ band largely overlaps with Johnson $B$ and SDSS $g$, while VT$\_R$ closely follows Cousins $R$ ($R_C$) and \textit{Gaia} $RP$, and is wider than the SDSS $r$ and $i$ filters.

To derive these transformations, we employed the ISAAC Spectroscopic Library provided by ESO,\footnote{\url{https://www.eso.org/sci/facilities/paranal/decommissioned/isaac/tools/lib.html}} which covers stars from early O to late M classes.
This ensures that the resulting relations are applicable over the main-sequence color range and for moderately evolved stars.
Synthetic photometry was performed using the {\sc python} package \texttt{synphot}~\citep{Lim2025_synphot}.
For each stellar spectrum, synthetic magnitudes were computed in the VT and external systems by convolving the flux distributions with the corresponding total system response curves.

To quantify the transformations, we modeled the magnitude differences for each comparison system as a function of appropriate color indices.
The resulting transformation relations were fitted using polynomials of the general form
\begin{equation}
    X - Y = \sum_{i=0}^{N}\,a_i\,C^i,
\end{equation}
where $X$ and $Y$ denote the magnitude in the VT or external photometric system, and $C$ is a color index sensitive to the stellar effective temperature (e.g., $B-V$, VT$\_B -$ VT$\_R$).
For transformations from external systems to VT, color indices native to the external system (e.g., $BP-RP$, $g-r$, $B-V$) were adopted.
For inverse transformations from VT to external systems, the color index was chosen as $\mathrm{VT}\_B - \mathrm{VT}\_R$, which provides a self-consistent and directly observable color within the VT system.
Here, $N$ denotes the maximum polynomial order.
In most cases, third-order polynomials ($N=3$) were adopted to capture nonlinear trends across the full color range, while lower-order ($N=2$) fits were used when they provided comparably good representations of the data.
Higher-order polynomials were tested but provided only marginal improvements while introducing instability at the color boundaries.
Therefore, third-order fits were selected as a balance between flexibility and robustness.
The best–fitting coefficients for all transformation relations are summarized in Table~\ref{tab:trans}, and the corresponding fitting curves with residuals are shown in Figures~\ref{fig:vt_trans_gaia}--\ref{fig:vt_trans_bessell}.

\begin{table*}[htbp]
\centering
\caption{Photometric transformation relations between the VT system and the \textit{Gaia}, Johnson--Cousins, and SDSS systems, derived from synthetic photometry of stellar spectra.
For readability, all listed coefficients and their uncertainties have been multiplied by $10^{2}$.
}
\label{tab:trans}
\setlength{\tabcolsep}{4pt}
\begin{tabular*}{\textwidth}{@{\extracolsep{\fill}} c c c
S[table-format=3.3,table-sign-mantissa]
S[table-format=1.3,table-sign-mantissa]
S[table-format=3.3,table-sign-mantissa]
S[table-format=1.3,table-sign-mantissa]
S[table-format=3.3,table-sign-mantissa]
S[table-format=1.3,table-sign-mantissa]
S[table-format=3.3,table-sign-mantissa]
S[table-format=1.3,table-sign-mantissa]
}
\hline\noalign{\smallskip}
$X$ & $Y$ & $C$
& \multicolumn{1}{c}{$a_0$} & \multicolumn{1}{c}{$a_{0,\,\rm err}$}
& \multicolumn{1}{c}{$a_1$} & \multicolumn{1}{c}{$a_{1,\,\rm err}$}
& \multicolumn{1}{c}{$a_2$} & \multicolumn{1}{c}{$a_{2,\,\rm err}$}
& \multicolumn{1}{c}{$a_3$} & \multicolumn{1}{c}{$a_{3,\,\rm err}$}\\
\hline\noalign{\smallskip}
\multicolumn{11}{c}{VT -- \textit{Gaia} Transformations} \\
\hline\noalign{\smallskip}
VT$\_B$ & $BP$ & $BP$ - $RP$ & -7.278 & 0.151 & -7.639 & 0.186 & 3.795 & 0.171 & -0.800 & 0.079 \\
VT$\_R$ & $RP$ & $BP$ - $RP$ & 1.586 & 0.035 & -1.246 & 0.044 & -1.570 & 0.040 & 0.405 & 0.018 \\
$BP$ & VT$\_B$ & VT$\_B$ - VT$\_R$ & 7.912 & 0.184 & 8.106 & 0.204 & -4.650 & 0.225 & 1.000 & 0.073 \\
$RP$ & VT$\_R$ & VT$\_B$ - VT$\_R$ & -1.430 & 0.040 & 1.379 & 0.045 & 1.591 & 0.051 & -0.422 & 0.016 \\
\hline\noalign{\smallskip}
\multicolumn{11}{c}{VT -- SDSS Transformations} \\
\hline\noalign{\smallskip}
VT$\_B$ & $g$ & $g$ - $r$ & 0.350 & 0.039 & -37.585 & 0.050 & -10.924 & 0.075 & \multicolumn{1}{c}{$\dots$} & \multicolumn{1}{c}{$\dots$} \\
VT$\_R$ & $r$ & $r$ - $i$ & 3.086 & 0.118 & -105.703 & 0.315 & -7.757 & 0.279 & \multicolumn{1}{c}{$\dots$} & \multicolumn{1}{c}{$\dots$} \\
VT$\_R$ & $z$ & $i$ - $z$ & 2.066 & 0.093 & 85.782 & 0.263 & -22.937 & 0.504 & \multicolumn{1}{c}{$\dots$} & \multicolumn{1}{c}{$\dots$} \\
$u$ & VT$\_B$ & VT$\_B$ - VT$\_R$ & 146.228 & 4.137 & 168.495 & 4.591 & -20.131 & 5.066 & -8.663 & 1.635 \\
$g$ & VT$\_B$ & VT$\_B$ - VT$\_R$ & 8.507 & 0.316 & 41.371 & 0.454 & 6.663 & 0.484 & -4.478 & 0.166 \\
$r$ & VT$\_R$ & VT$\_B$ - VT$\_R$ & -13.128 & 0.370 & 47.215 & 0.439 & 10.510 & 0.487 & 0.992 & 0.163 \\
$i$ & VT$\_R$ & VT$\_B$ - VT$\_R$ & -4.202 & 0.087 & 2.398 & 0.116 & 3.633 & 0.121 & 0.484 & 0.051 \\
$z$ & VT$\_R$ & VT$\_B$ - VT$\_R$ & 10.511 & 0.385 & -26.352 & 0.427 & 0.549 & 0.472 & -0.468 & 0.152 \\
\hline\noalign{\smallskip}
\multicolumn{11}{c}{VT -- Johnson--Cousins Transformations} \\
\hline\noalign{\smallskip}
VT$\_B$ & $B$ & $B$ - $V$ & 1.019 & 0.067 & -70.821 & 0.118 & -11.841 & 0.157 & \multicolumn{1}{c}{$\dots$} & \multicolumn{1}{c}{$\dots$} \\
VT$\_R$ & $R_C$ & $R_C$ - $I_C$ & 0.933 & 0.077 & -77.310 & 0.189 & -7.574 & 0.234 & \multicolumn{1}{c}{$\dots$} & \multicolumn{1}{c}{$\dots$} \\
$U$ & VT$\_B$ & VT$\_B$ - VT$\_R$ & 142.971 & 4.079 & 171.401 & 4.526 & -18.269 & 4.994 & -9.298 & 1.612 \\
$B$ & VT$\_B$ & VT$\_B$ - VT$\_R$ & 25.985 & 0.837 & 66.297 & 0.980 & -0.348 & 1.088 & -4.186 & 0.348 \\
$V$ & VT$\_B$ & VT$\_B$ - VT$\_R$ & -8.811 & 0.204 & -17.627 & 0.225 & 8.035 & 0.249 & -0.646 & 0.080 \\
$R_C$ & VT$\_R$ & VT$\_B$ - VT$\_R$ & -10.780 & 0.239 & 35.570 & 0.300 & 7.204 & 0.136 & \multicolumn{1}{c}{$\dots$} & \multicolumn{1}{c}{$\dots$}\\
$I_C$ & VT$\_R$ & VT$\_B$ - VT$\_R$ & 1.995 & 0.243 & -12.737 & 0.271 & 0.251 & 0.304 & 1.082 & 0.097 \\
\hline\noalign{\smallskip}
\end{tabular*}
\end{table*}

\subsubsection{Transformation between the VT and \textit{Gaia}~DR3 Systems}
\label{subsubsec:tran_gaia}
A comparison of the response curves (Fig.~\ref{fig:cmp_filters}) shows that VT$\_R$ closely resembles the \textit{Gaia} $RP$ band, whereas VT$\_B$ lies between \textit{Gaia} $BP$ and SDSS $g$.
These similarities suggest that relatively simple color-dependent transformations may be expected, particularly for VT$\_R$.

\begin{figure*}
\centering
\includegraphics[width=0.24\textwidth, angle=0]{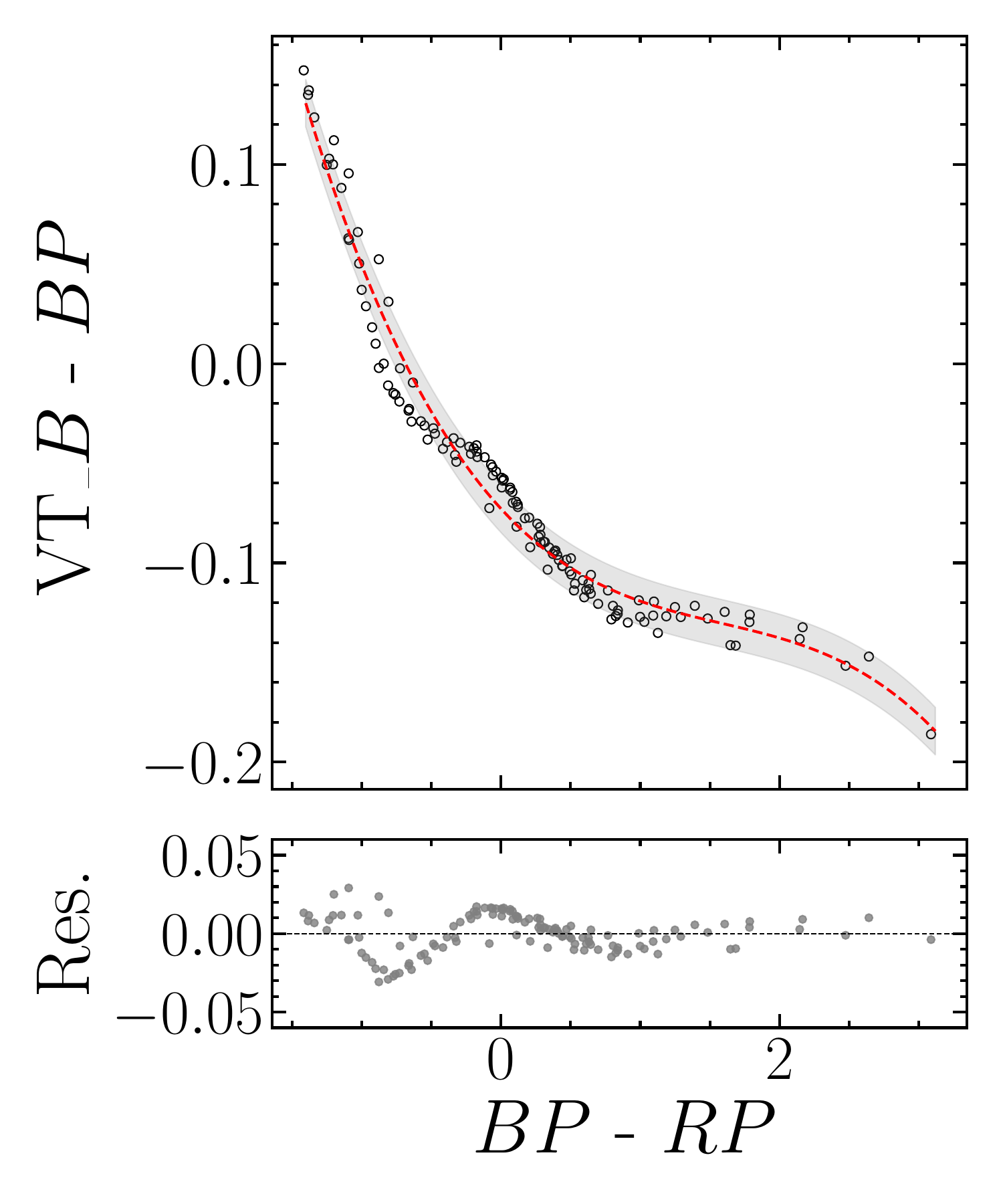}
\includegraphics[width=0.24\textwidth, angle=0]{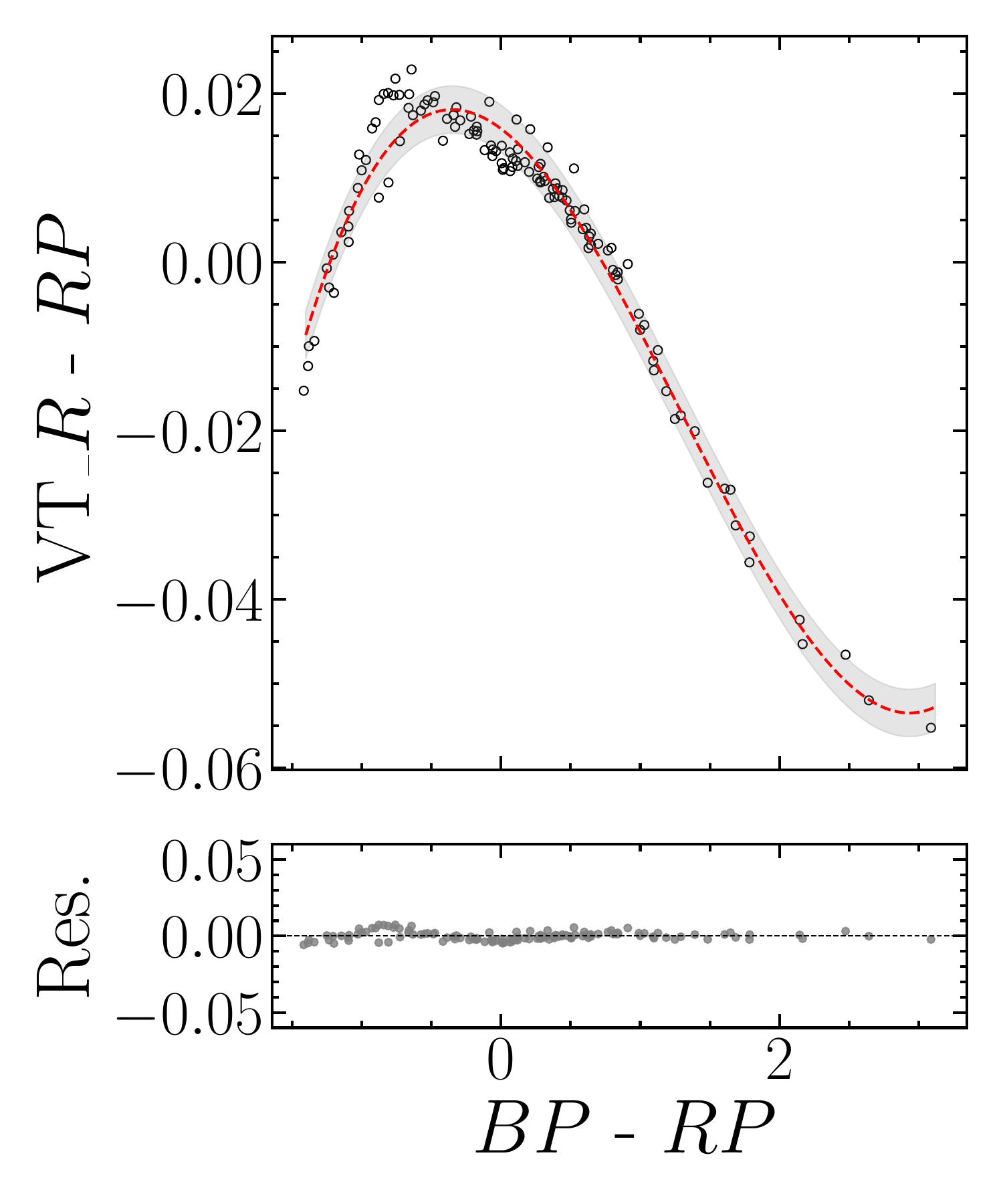}
\includegraphics[width=0.24\textwidth, angle=0]{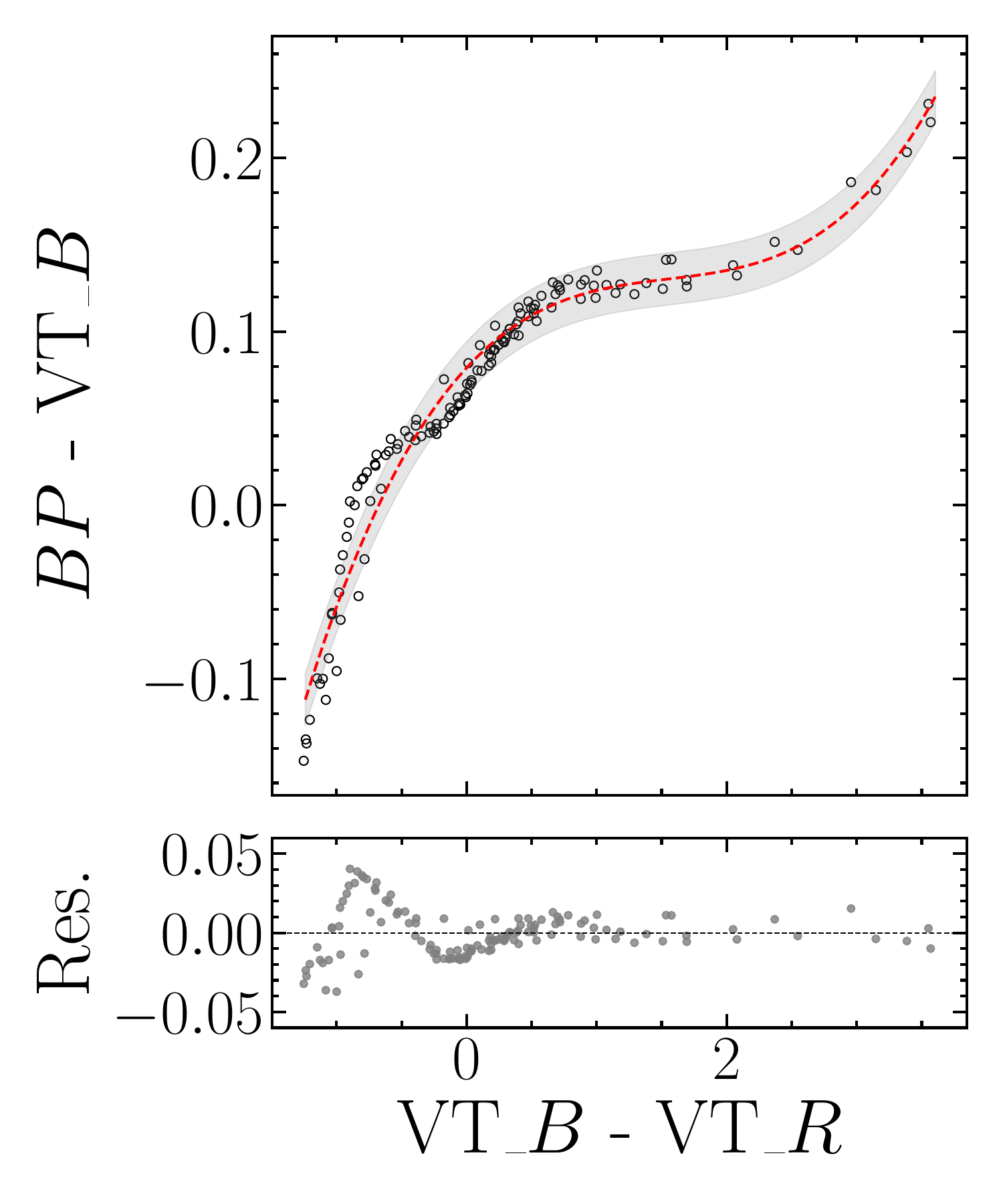}
\includegraphics[width=0.24\textwidth, angle=0]{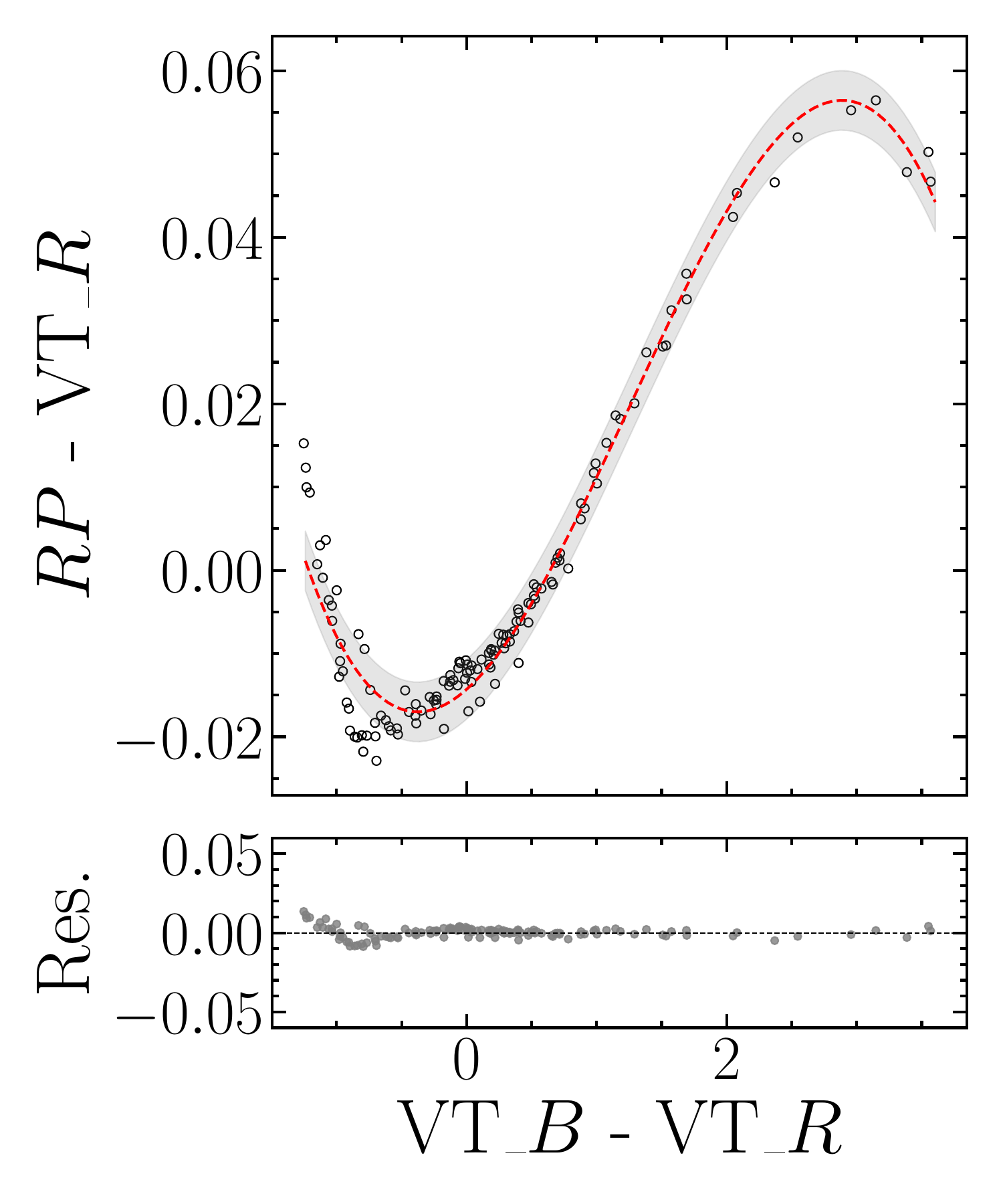}
\caption{Magnitude differences between the VT bands and the \textit{Gaia} photometric system as a function of color indices, derived from synthetic photometry of stellar spectra.
In each panel, black circles represent the synthetic photometric data points, the red dashed line shows the best--fit transformation relation, and the gray shaded region indicates the $1\sigma$ uncertainty of the fit.
The lower subpanels display the residuals relative to the best--fit relation.
}
\label{fig:vt_trans_gaia}
\end{figure*}

We derived bidirectional transformation relations between the VT and \textit{Gaia}~DR3 systems~\citep{Riello2021} using synthetic photometry of stellar spectra. 
For transformations from \textit{Gaia} to VT, the color index $(BP-RP)$ was adopted, while for inverse transformations from VT to \textit{Gaia}, the color index $\mathrm{VT}\_B - \mathrm{VT}\_R$ was used to ensure internal consistency within the VT system (Table~\ref{tab:trans}). The results are shown in Figure~\ref{fig:vt_trans_gaia}.

For VT$\_R$, the quantity $\mathrm{VT}\_R - RP$ as a function of $(BP-RP)$ exhibits only weak color dependence over the range $-1.5 \lesssim (BP-RP) \lesssim 3$. 
A third-order polynomial was adopted to account for subtle nonlinear trends across the full color range. 
The residual scatter remains at the $\sim0.01$~mag level for most spectral types, indicating that VT$\_R$ is closely matched to \textit{Gaia} $RP$, with only minor color-dependent corrections required.

The transformation between VT$\_B$ and \textit{Gaia} $BP$ shows more pronounced curvature.
Because VT$\_B$ is narrower and slightly redder than $BP$, the difference $\mathrm{VT}\_B - BP$ varies systematically with stellar color.
This relation is well described by a third-order polynomial, with residuals typically within $\sim0.03$~mag across the main-sequence color range.
The remaining low-level structure at the $\sim0.01$~mag level likely reflects differences in detailed passband shapes and the influence of spectral features.

In addition, all the \textit{Gaia}-based transformation relations show noticeably larger dispersion toward the blue end, and stars with $(BP-RP)\approx-1$ should be treated with particular caution, as the fits become less reliable in this regime due to the limited number of calibration spectra and the increasing mismatch between the VT and \textit{Gaia} passbands.
The inverse transformations ($BP$ and $RP$ expressed as functions of $\mathrm{VT}\_B - \mathrm{VT}\_R$) exhibit comparable residual levels, confirming the internal consistency of the bidirectional fitting.

\subsubsection{Transformation between the VT and SDSS Systems}
\label{subsubsec:tran_sdss}
\begin{figure*}
\raggedright
\includegraphics[width=0.19\textwidth]{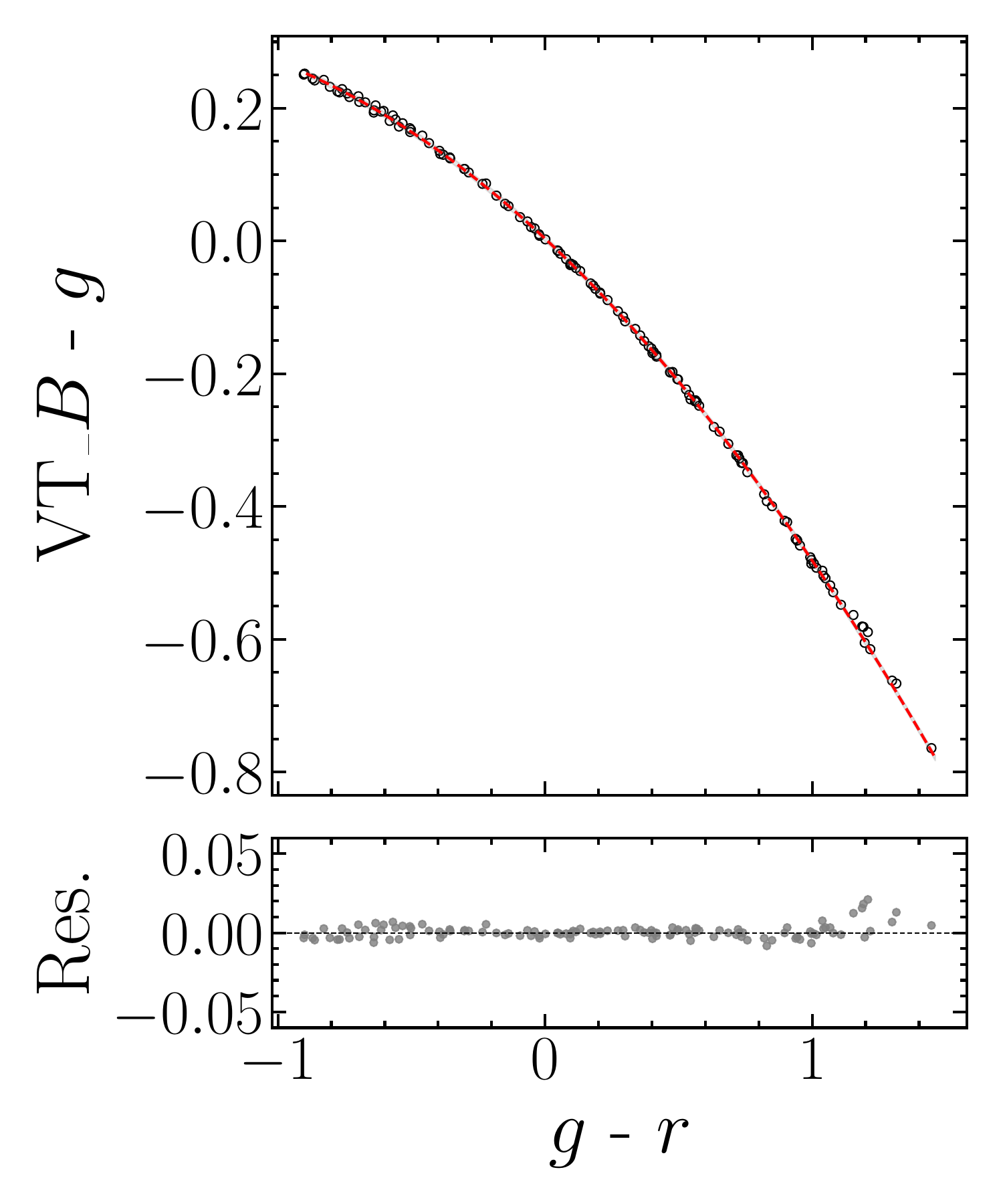}\hspace{0.005\textwidth}
\includegraphics[width=0.19\textwidth]{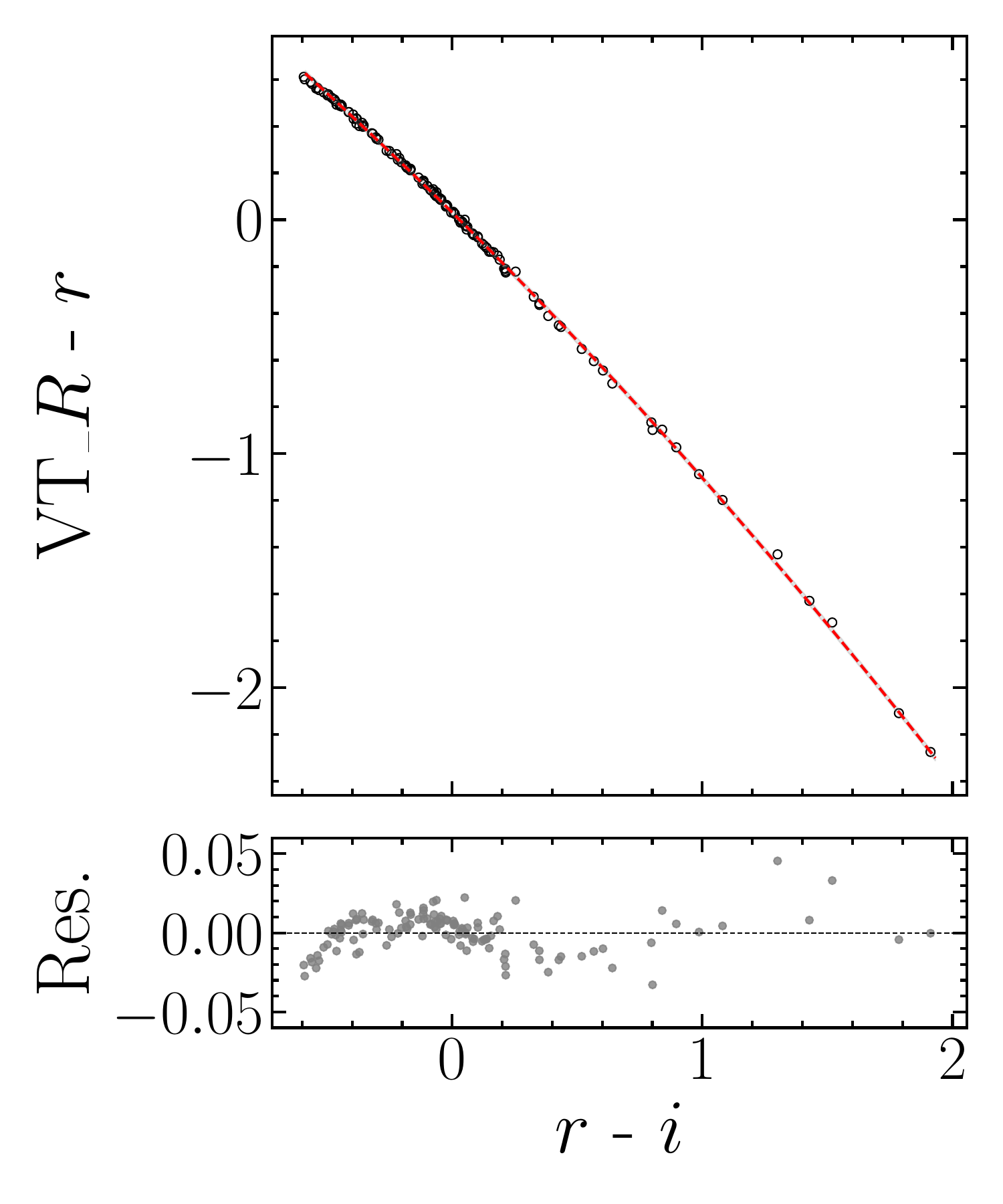}\hspace{0.005\textwidth}
\includegraphics[width=0.19\textwidth]{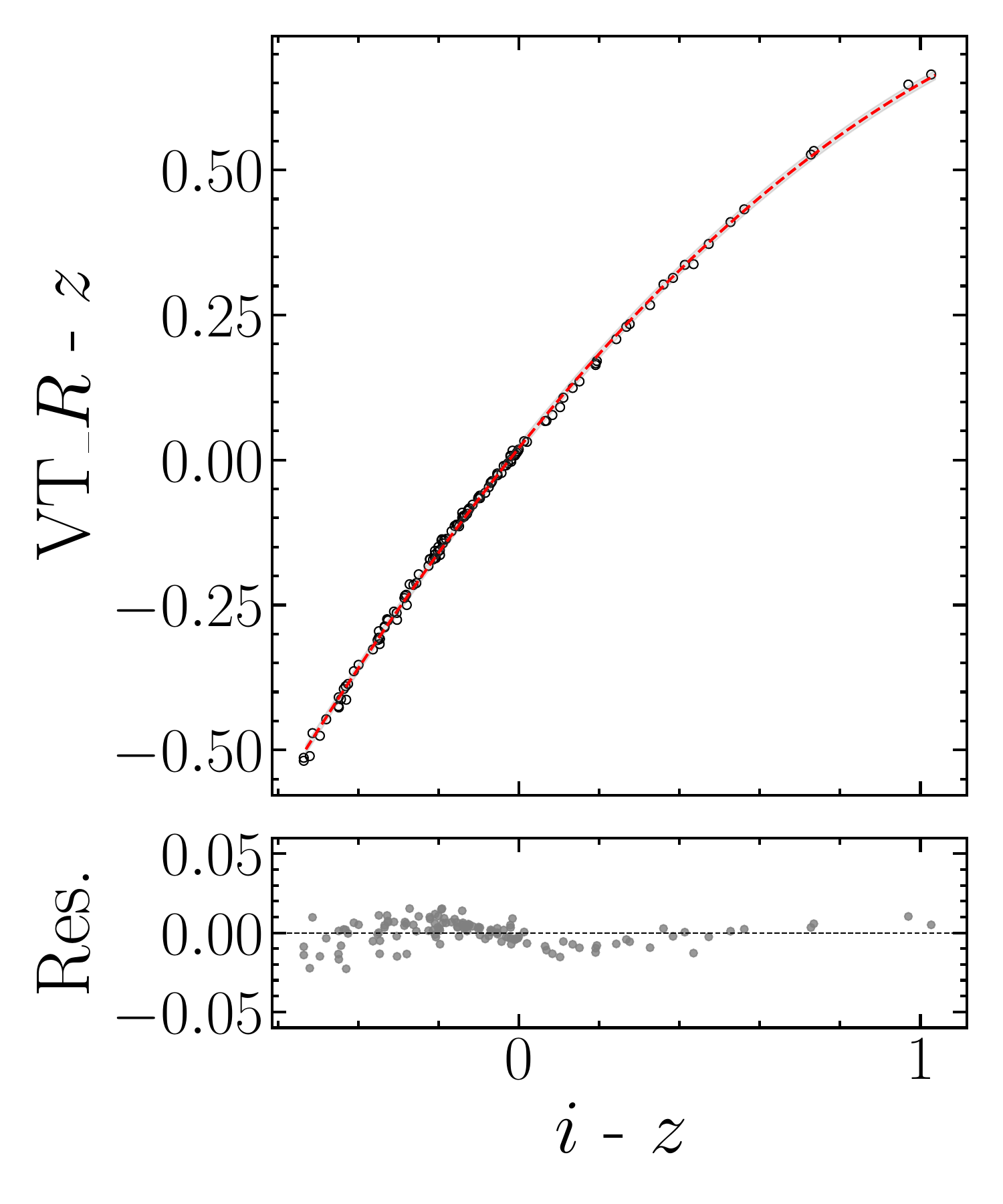}
\par\vspace{2mm}
\includegraphics[width=0.19\textwidth]{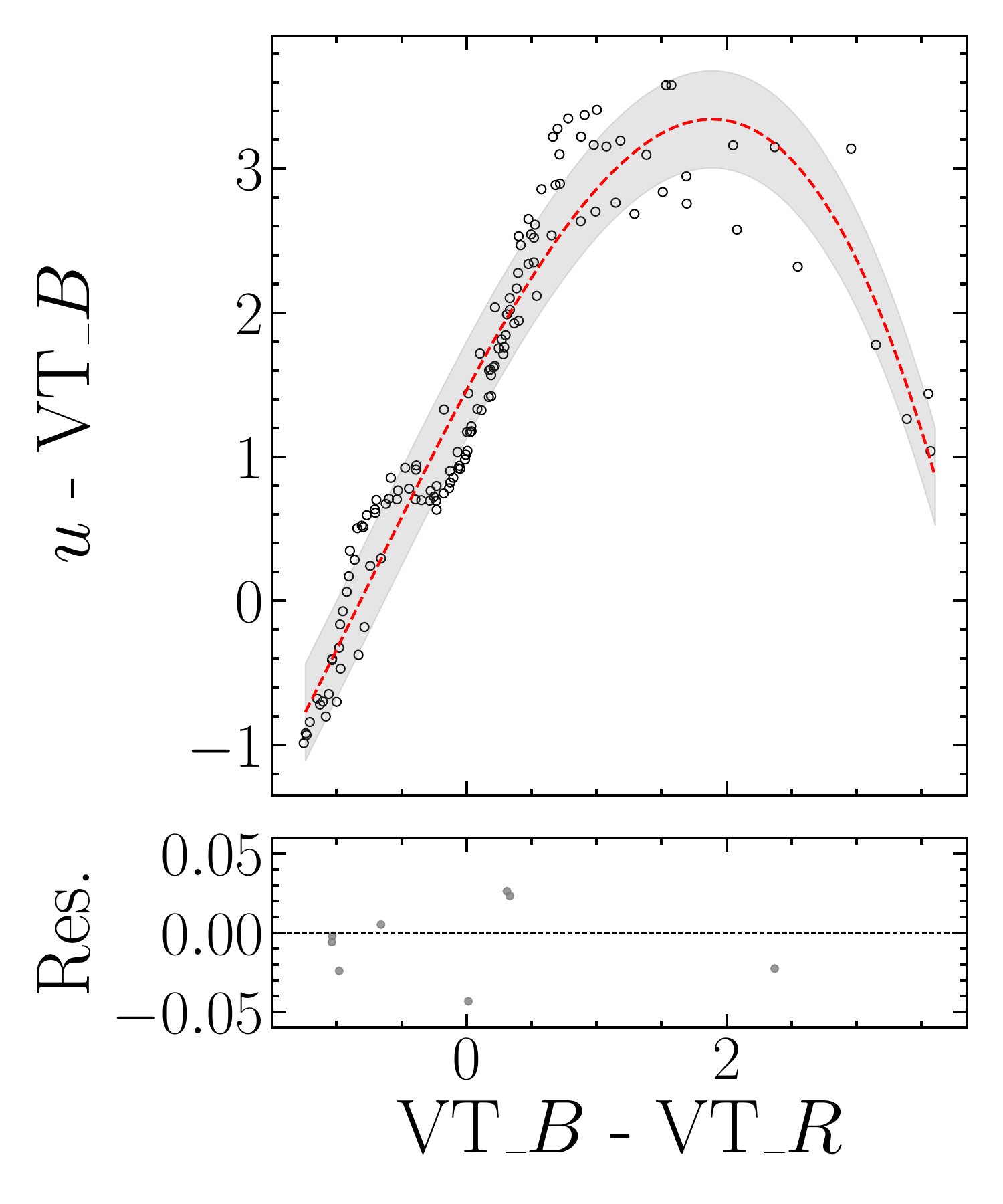}\hspace{0.005\textwidth}
\includegraphics[width=0.19\textwidth]{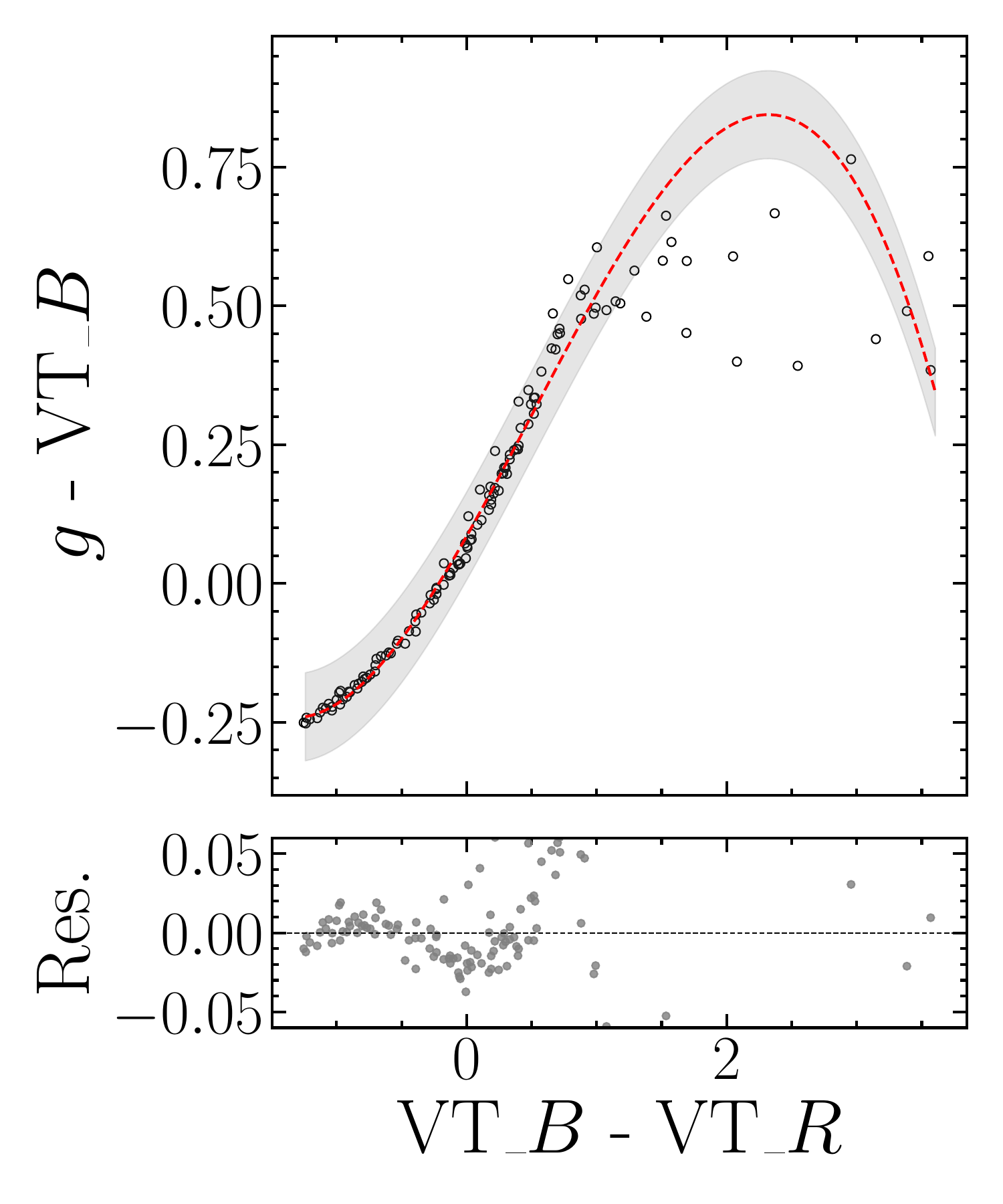}\hspace{0.005\textwidth}
\includegraphics[width=0.19\textwidth]{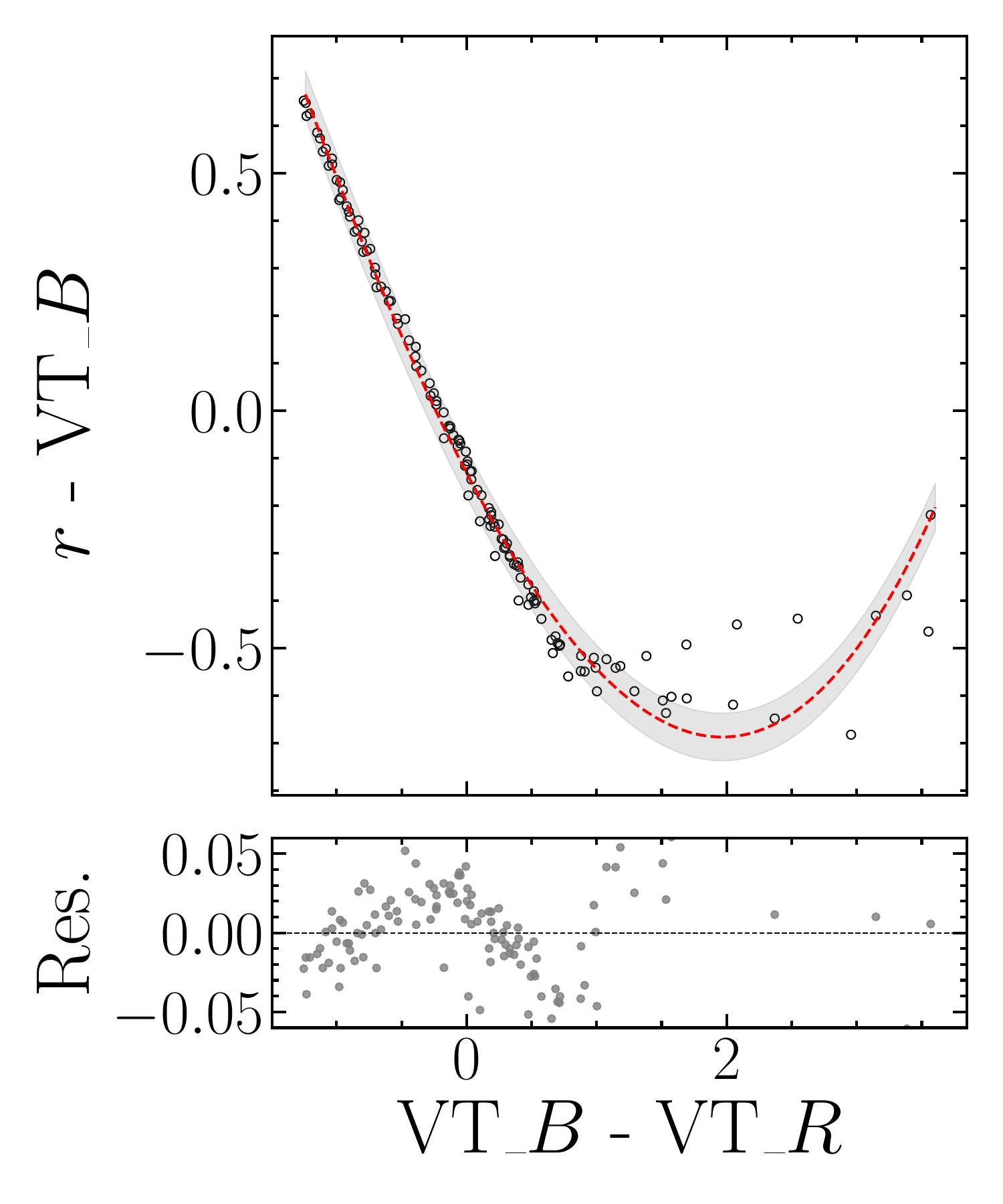}\hspace{0.005\textwidth}
\includegraphics[width=0.19\textwidth]{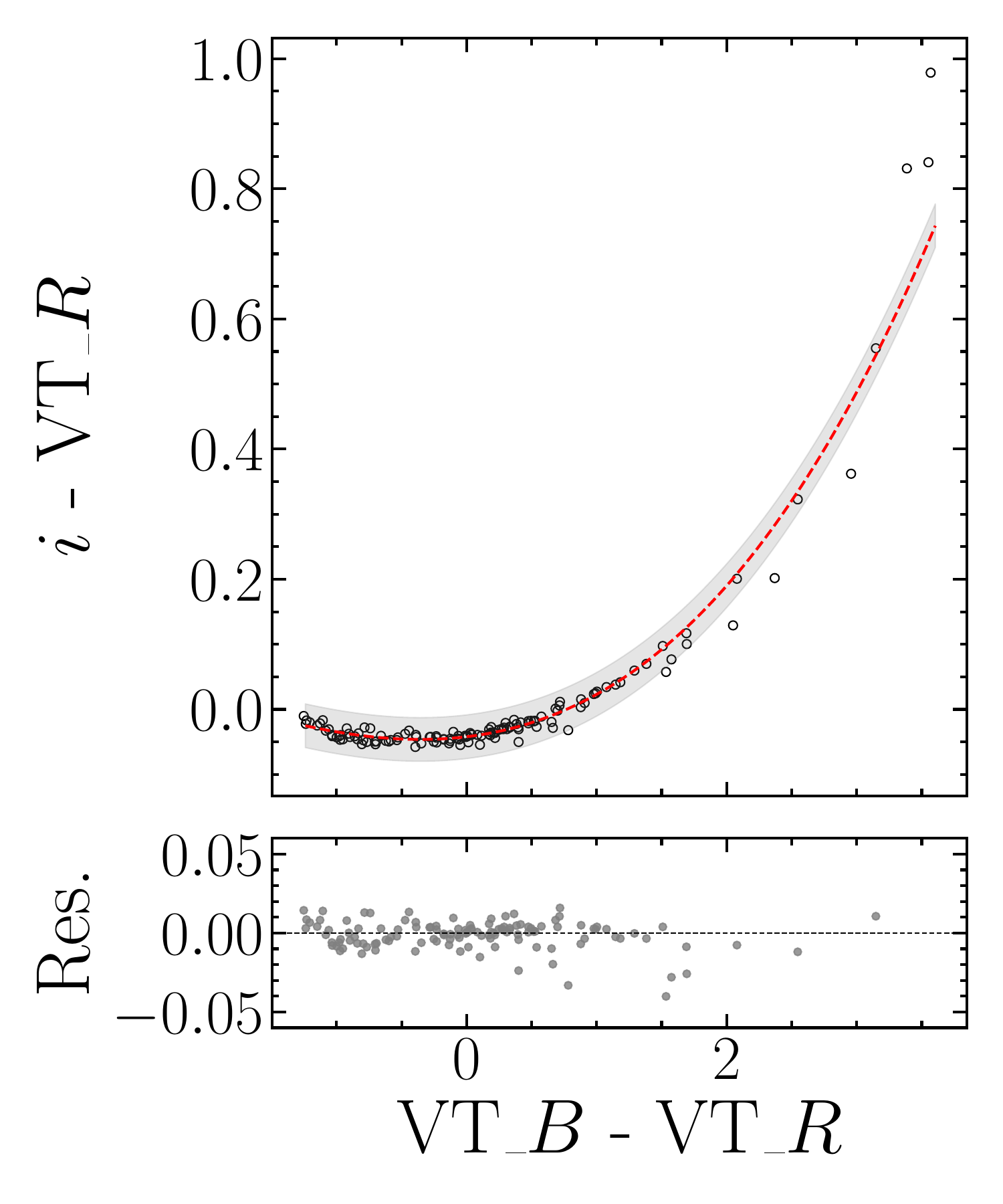}\hspace{0.005\textwidth}
\includegraphics[width=0.19\textwidth]{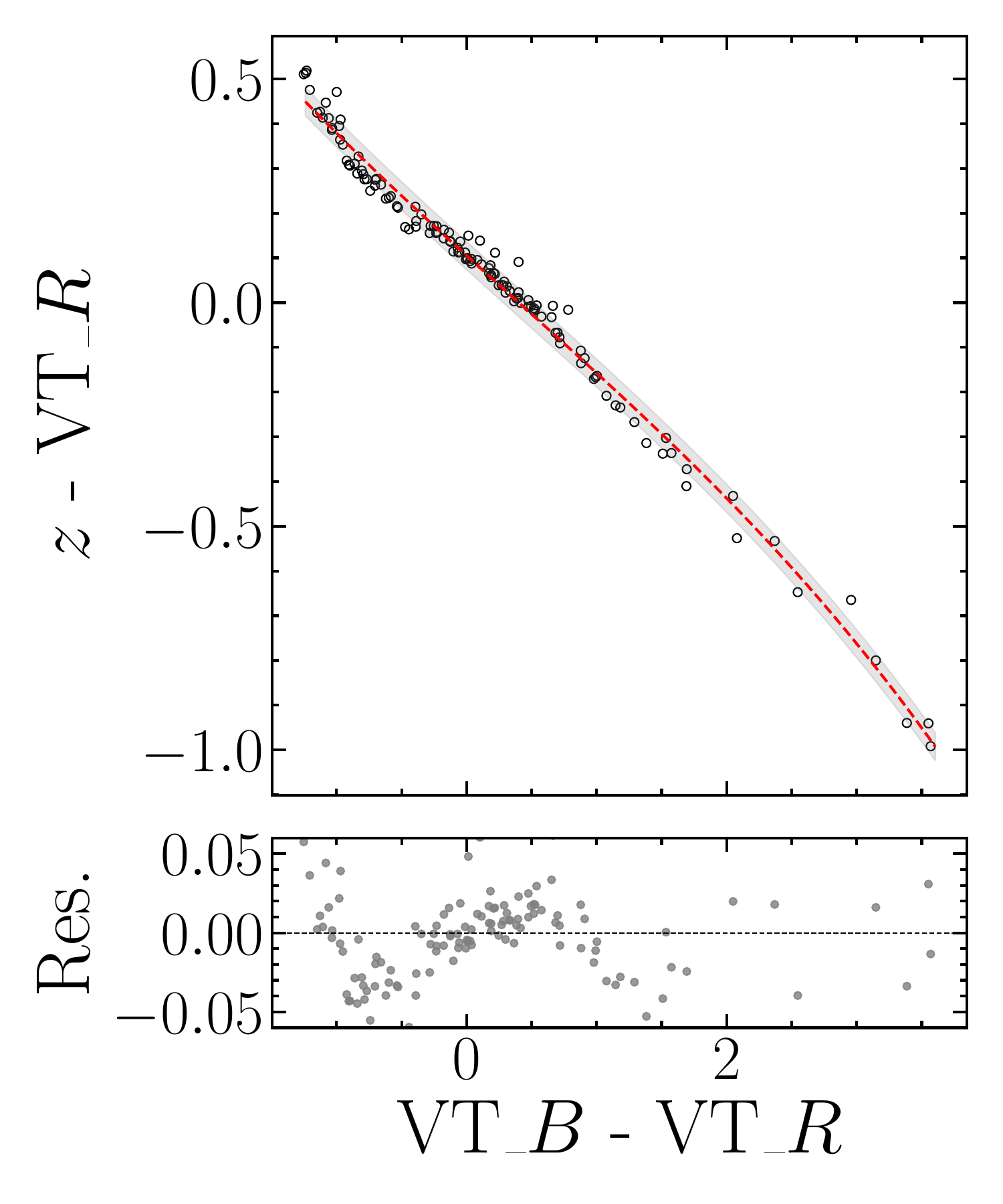}
\caption{Magnitude differences between the VT bands and the SDSS photometric system as a function of color indices, derived from synthetic photometry of stellar spectra.
The symbols, lines, and shaded regions have the same meaning as in Fig.~\ref{fig:vt_trans_gaia}.}
\label{fig:vt_trans_sdss}
\end{figure*}

Figure~\ref{fig:vt_trans_sdss} presents the bidirectional transformation relations between the VT and SDSS systems~\citep{Doi2010}, derived from synthetic stellar photometry.
The upper panels show the transformations from SDSS to VT, while the lower panels display the inverse relations.
For transformations from SDSS to VT, native SDSS color indices (e.g., $g-r$, $r-i$, $i-z$) were adopted. 
For inverse transformations from VT to SDSS, the internal VT color $\mathrm{VT}\_B - \mathrm{VT}\_R$ was used to ensure practical applicability when only VT measurements are available (Table~\ref{tab:trans}).

For VT$\_B$, the most relevant SDSS comparisons involve the $g$ band. 
The relations $\mathrm{VT}\_B - g$ and $\mathrm{VT}\_B - r$ as functions of $(g-r)$ show clear nonlinear trends and are therefore modeled with third-order polynomials. 
We tested higher-order fits (fourth order and above), which yield marginal reductions in RMS scatter but introduce increased oscillatory behavior at the color boundaries without significant improvement in predictive accuracy. 
A cubic formulation was therefore adopted as a compromise between flexibility and stability.

The VT$\_R$ band most closely resembles the SDSS $r$ and partially overlaps with $z$. 
The relations $\mathrm{VT}\_R - r$ as a function of $(r-i)$ and $\mathrm{VT}\_R - z$ as a function of $(i-z)$ exhibit smooth curvature and are well represented by quadratic polynomials over the color ranges $-0.5 \lesssim (r-i) \lesssim 2.0$ and $-0.5 \lesssim (i-z) \lesssim 2.0$. 
Typical residuals remain within $\sim0.03$~mag across the main-sequence regime. 
A moderate increase in scatter is observed toward the reddest colors, which is primarily attributable to the reduced number of late-type calibration spectra and the increasing sensitivity of red passbands to molecular absorption features.

The inverse relations (e.g., $g$, $r$, $i$, $z$ expressed as functions of $\mathrm{VT}\_B - \mathrm{VT}\_R$) show comparable residual levels and confirm the internal consistency of the bidirectional fitting procedure.
For very red stars ($\mathrm{VT}\_B - \mathrm{VT}\_R \gtrsim 1.0$), the transformation relations show noticeably larger dispersion and deviate from a simple polynomial description.
This behavior is primarily driven by late-type (M-type) stars, for which broadband colors become increasingly degenerate due to strong molecular absorption features and the limited sensitivity of $\mathrm{VT}\_B - \mathrm{VT}\_R$ to spectral subtype.
The derived relations should therefore be applied with caution in this color regime.

Overall, within the primary stellar color range $-1.0 \lesssim (g-r) \lesssim 1.0$, the VT magnitudes can be transformed into the SDSS system with typical uncertainties of $\sim0.03$~mag. 
The relations remain stable and monotonic over this interval, with no evidence of strong pathological behavior.
In contrast, applications involving $\mathrm{VT}\_B - \mathrm{VT}\_R \gtrsim 1.0$ should be treated with caution, as the transformation relations become increasingly uncertain in this regime.

\subsubsection{Transformation between the VT and Johnson--Cousins Systems}
\label{subsubsec:tran_jc}
The Johnson--Cousins photometric system remains widely used in the literature, particularly for long-term stellar monitoring and historical photometric data.
We therefore derived transformation relations between the VT system and the standard Johnson--Cousins $UBV(RI)_C$ system~\citep{Bessell1990}.

For transformations from Johnson--Cousins to VT, the color indices $(B-V)$ and $(R_C-I_C)$ were adopted for the VT$\_B$ and VT$\_R$ relations, respectively.
For the inverse transformations, the internal VT color $\mathrm{VT}\_B - \mathrm{VT}\_R$ was used to express $U$, $B$, $V$, $R_C$, and $I_C$ magnitudes (Table~\ref{tab:trans}).

The relation $\mathrm{VT}\_B - B$ as a function of $(B-V)$ exhibits smooth nonlinear behavior and is well represented by a low-order polynomial across the color range $-0.9 \lesssim (B-V) \lesssim 1.5$. 
Typical residuals remain within $\sim0.03$~mag for most main-sequence and moderately evolved stars. 
Similarly, $\mathrm{VT}\_R - R_C$ as a function of $(R_C-I_C)$ shows stable behavior over $-0.9 \lesssim (R_C-I_C) \lesssim 1.5$, with comparable residual levels.

At the edges of the calibration range, some increased dispersion is observed.
For the VT$\_B$ transformation, red stars with $(B-V)\gtrsim0.5$ exhibit larger scatter, while for the VT$\_R$ transformation the bluest stars with $(R_C-I_C)\lesssim-0.5$ show a noticeable increase in residuals.
In particular, for very red stars with $\mathrm{VT}_B - \mathrm{VT}_R \gtrsim 1.0$ (as discussed in Sect~\ref{subsubsec:tran_sdss}), the polynomial representation becomes less accurate and the scatter increases noticeably.
Applications in this color regime should therefore be treated with caution.
Despite these edge effects, the transformations remain stable and well behaved within the primary stellar color ranges, with typical uncertainties of $\sim0.03$~mag.

Because the Johnson--Cousins system is traditionally tied to the Vega magnitude scale, whereas the SDSS systems are defined on an AB-like scale, some care is required when combining measurements across different photometric systems.
However, our synthetic-photometry approach treats all systems in a fully consistent manner, such that differences in absolute zero point definitions are naturally absorbed into the fitted coefficients.
The resulting transformations therefore implicitly include the effective zero point offsets between the VT and Johnson--Cousins magnitudes implied by the adopted flux calibration.

\begin{figure*}
\raggedright
\includegraphics[width=0.19\textwidth, angle=0]{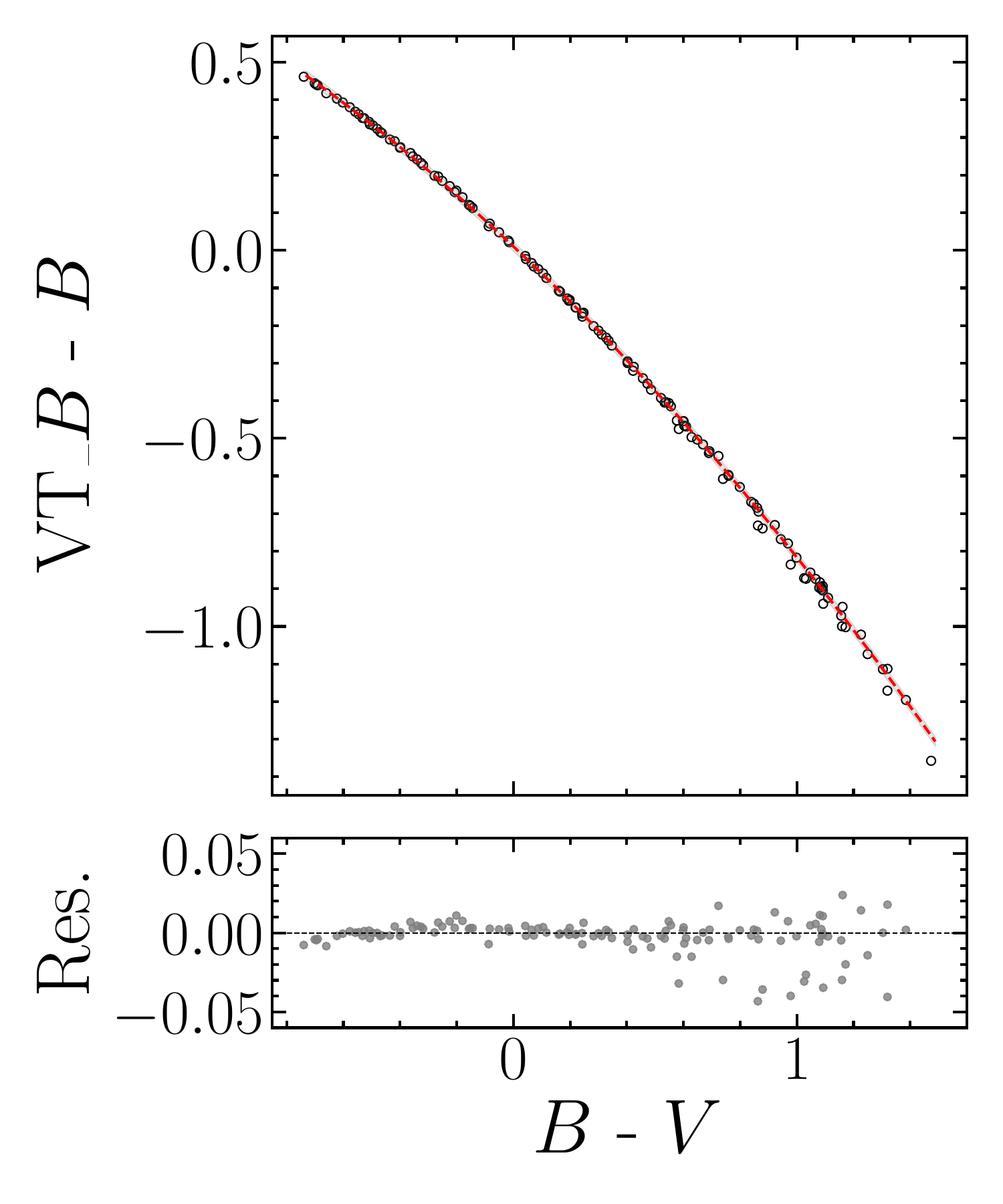}
\includegraphics[width=0.19\textwidth, angle=0]{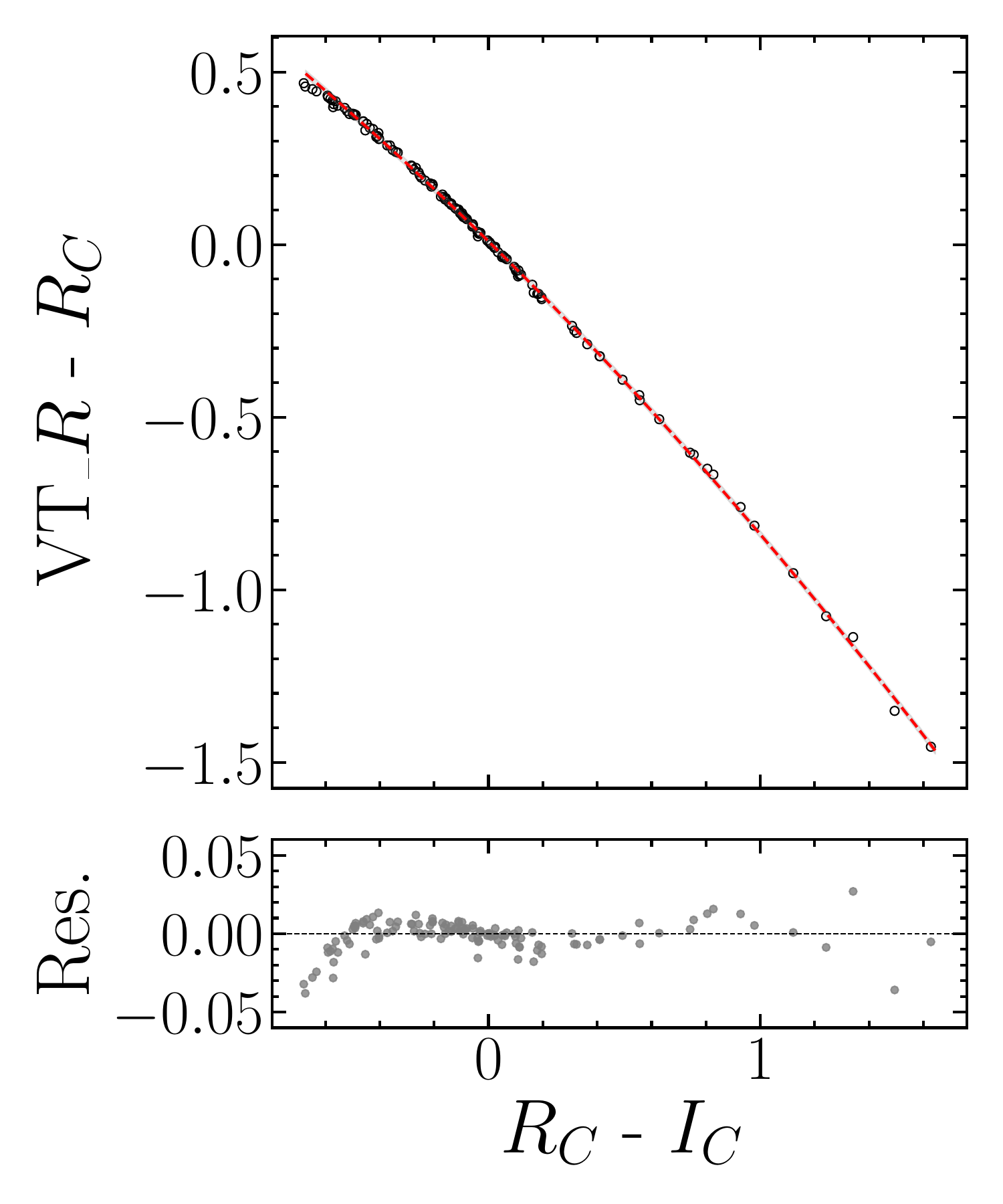}
\par\vspace{2mm}
\includegraphics[width=0.19\textwidth, angle=0]{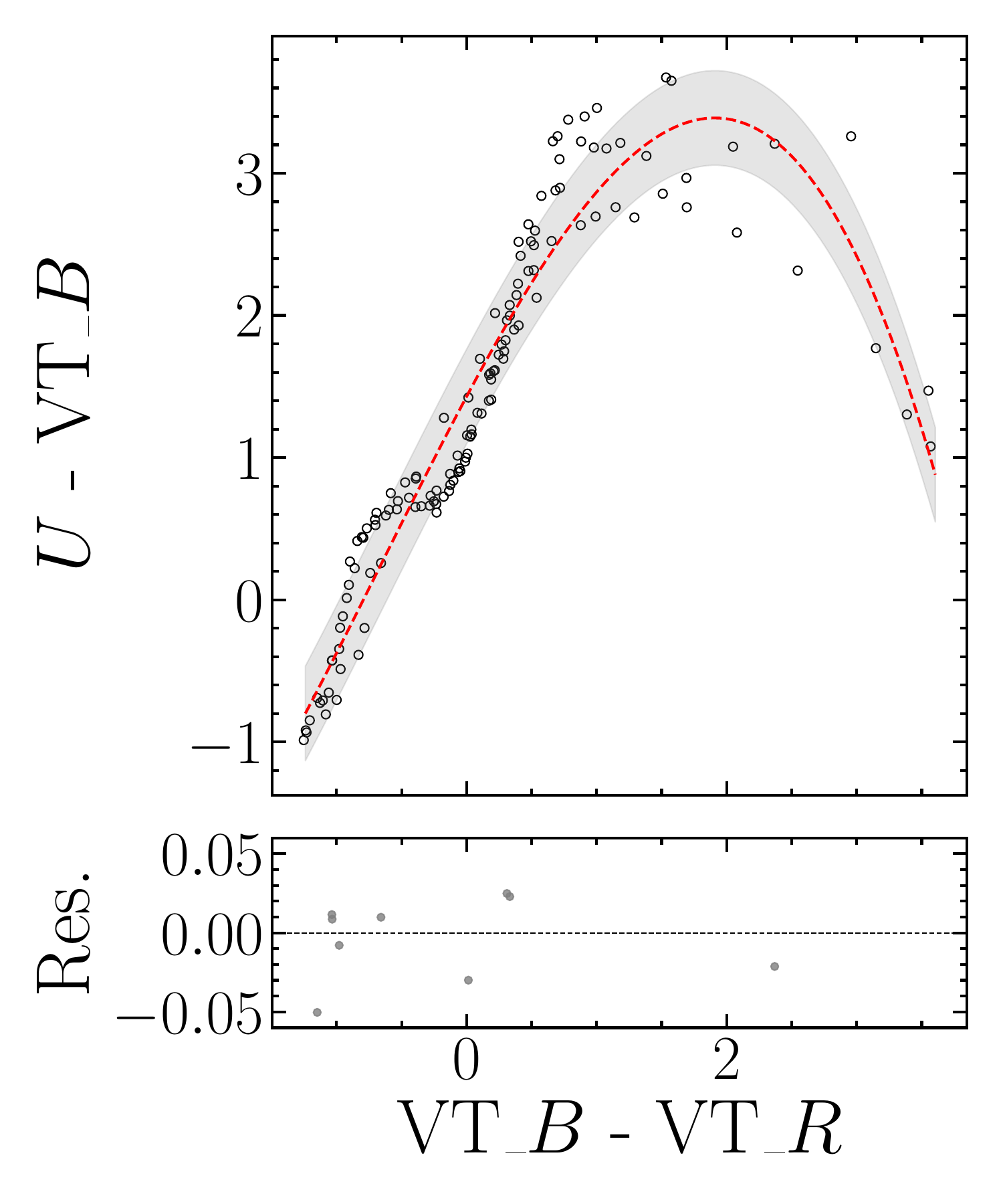}
\includegraphics[width=0.19\textwidth, angle=0]{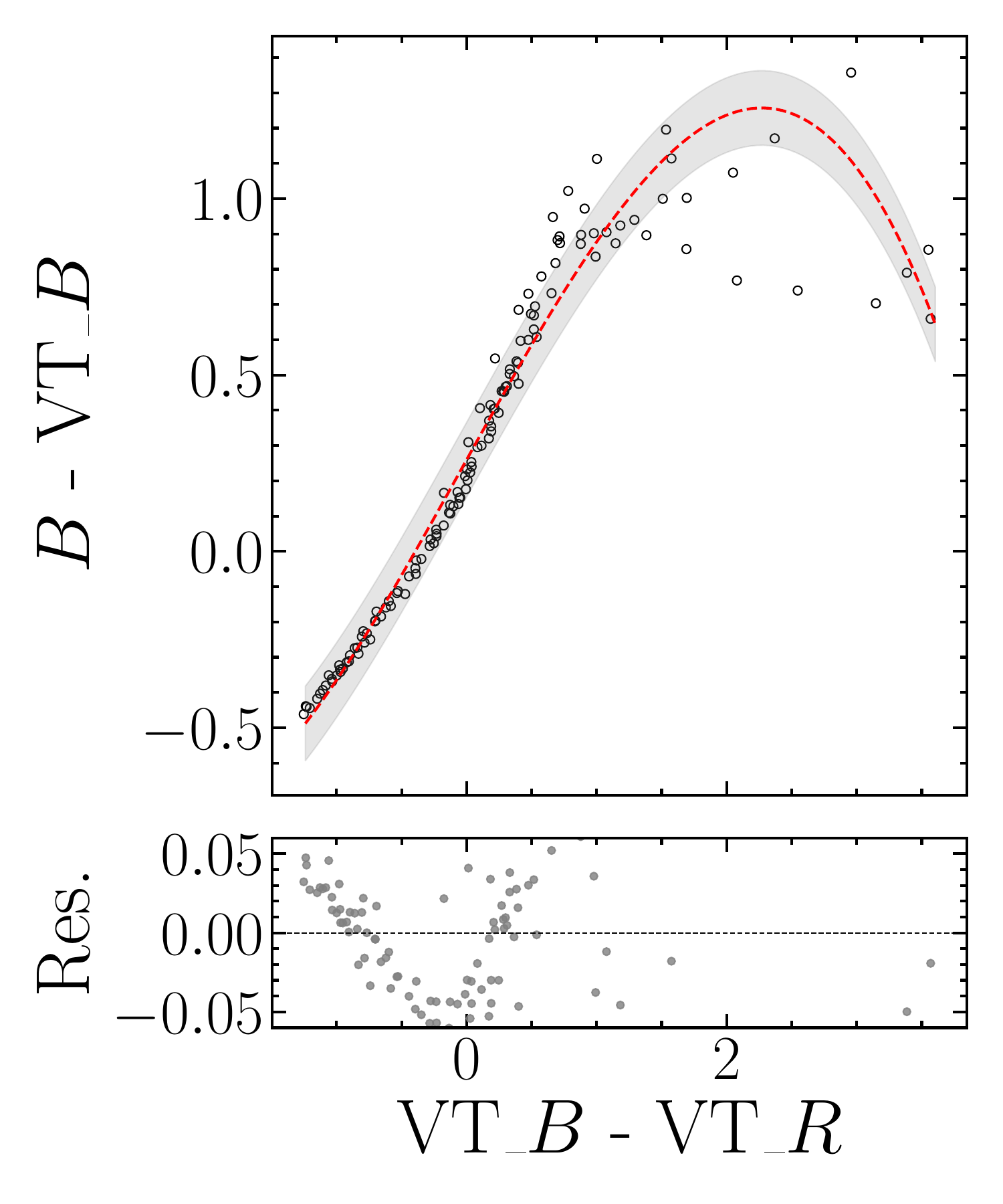}
\includegraphics[width=0.19\textwidth, angle=0]{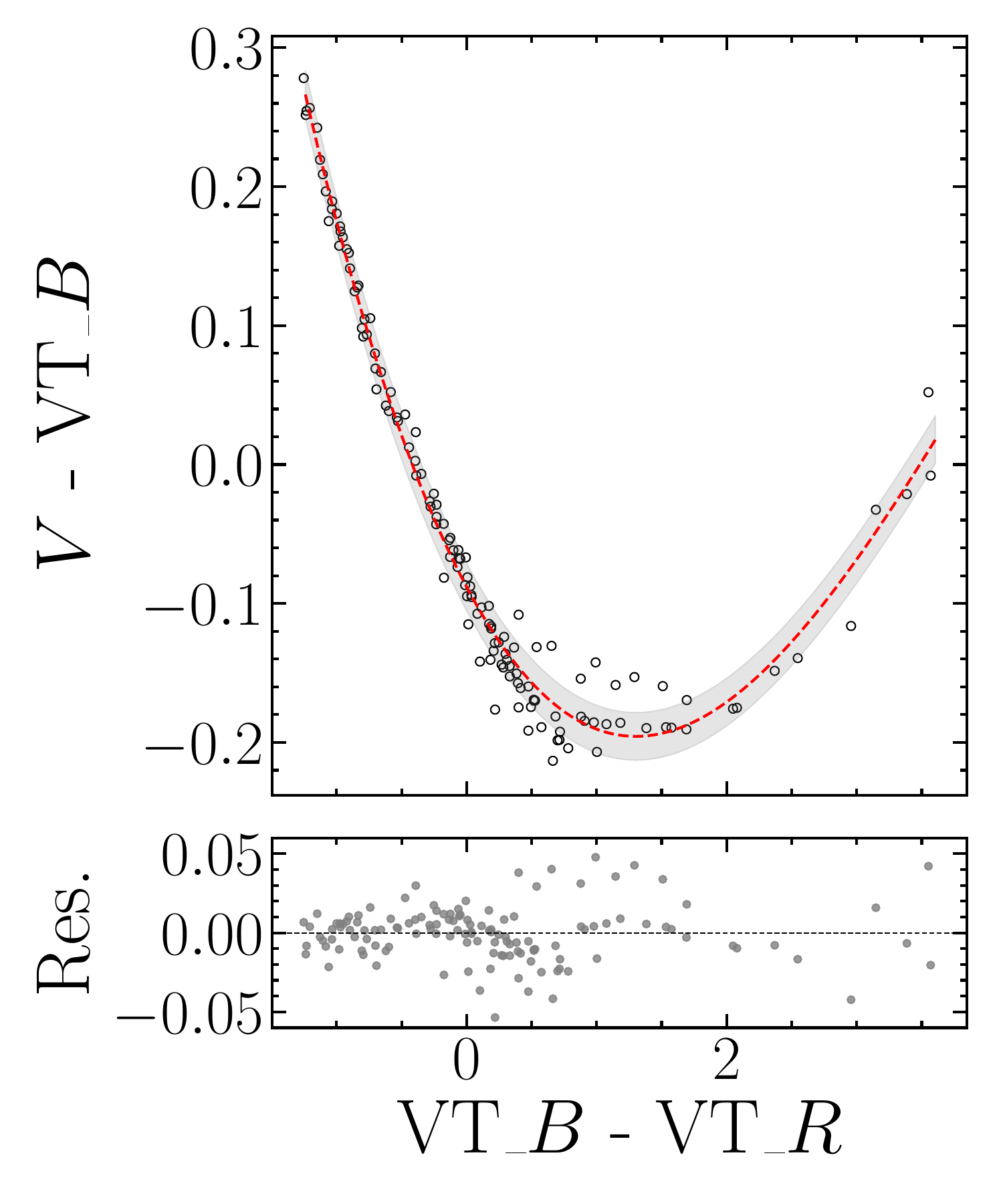}
\includegraphics[width=0.19\textwidth, angle=0]{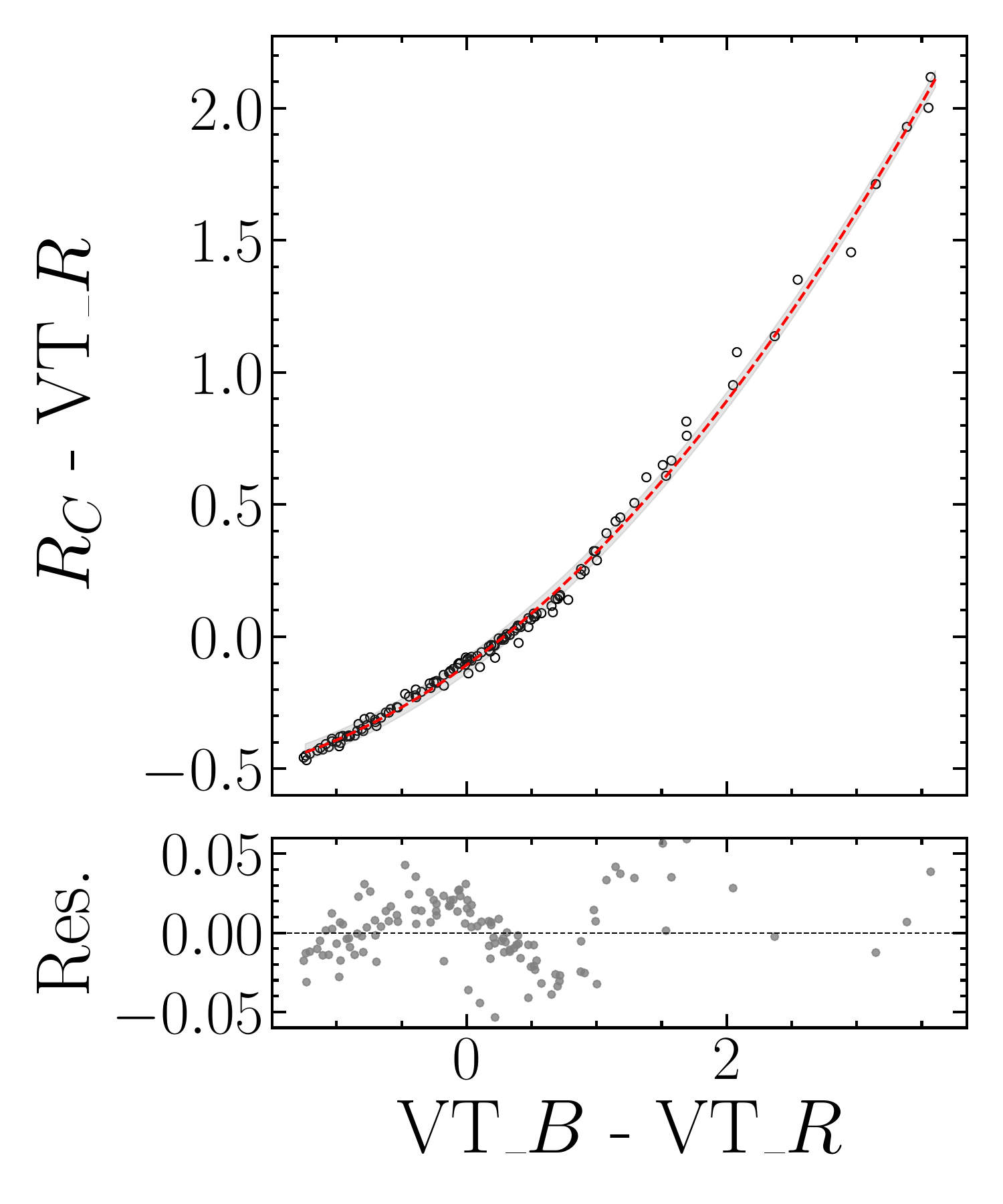}
\includegraphics[width=0.19\textwidth, angle=0]{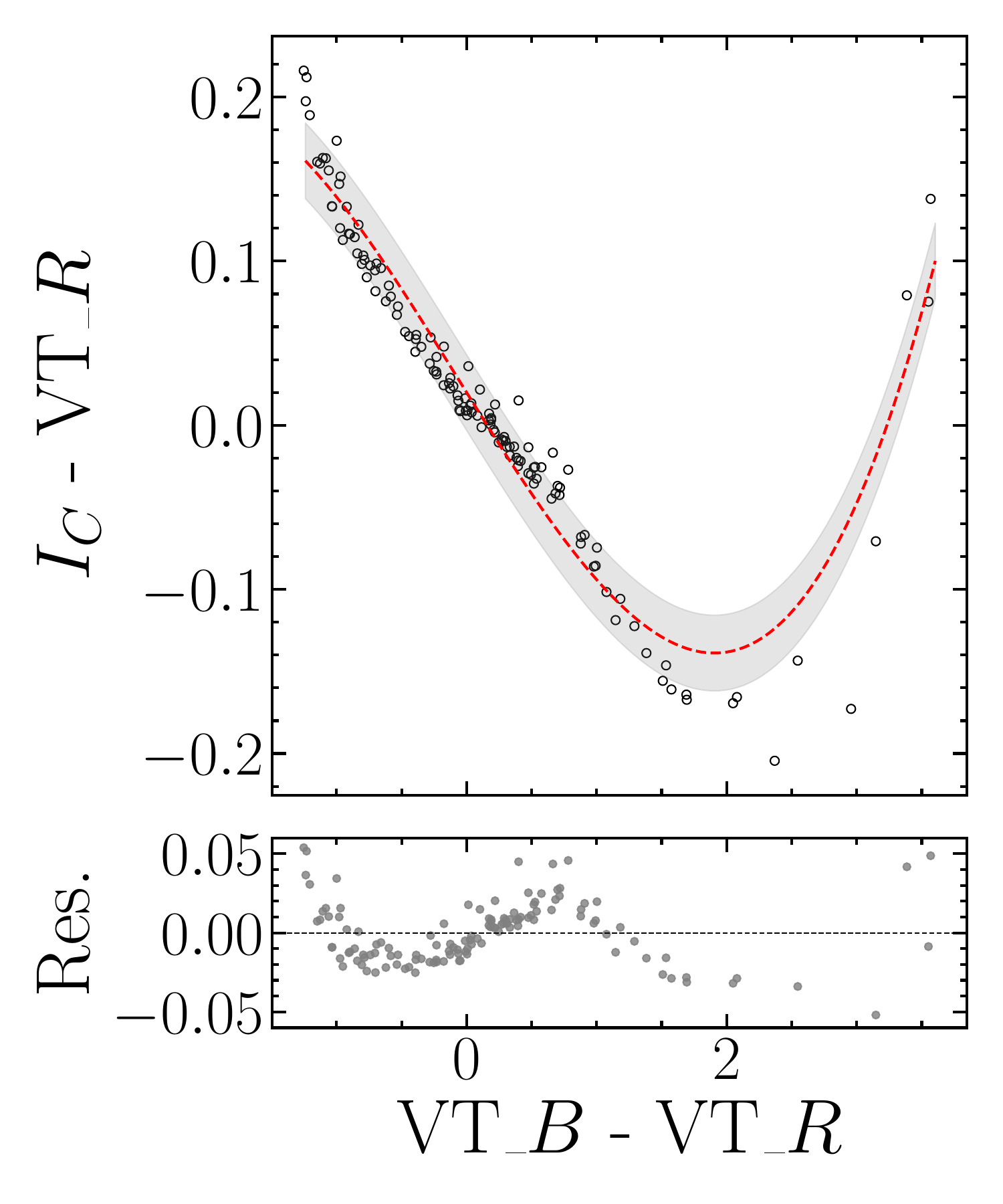}
\caption{Magnitude differences between the VT bands and the Johnson--Cousins photometric system as a function of color indices, derived from synthetic photometry of stellar spectra.
The symbols, lines, and shaded regions have the same meaning as in Fig.~\ref{fig:vt_trans_gaia}.}
\label{fig:vt_trans_bessell}
\end{figure*}

\section{Discussion}
\label{sec:dis}

\subsection{Contamination and Effects on Photometric Zero Points}
\label{subsec:contamination}

As described in Section~\ref{subsec:stds}, the long-term monitoring of spectrophotometric standard stars reveals significant temporal variations in the VT photometric zero point.
Shortly after launch, the zero point exhibited a rapid and substantial decline, which cannot be attributed to intrinsic variability of the standard stars or to statistical measurement uncertainties,  but instead points to an instrumental origin.

A natural explanation for these behaviors is optical contamination within the VT system.
For space-borne instruments, outgassed material (e.g., $\mathrm{H_{2}O}$ molecules) from the spacecraft platform and onboard components can deposit on sensitive optical surfaces.
At the typically low operating temperatures, such contaminants may accumulate and undergo photochemical processing, leading to a gradual reduction in system throughput.
In addition, contaminant-induced scattering can introduce stray light and short-term fluctuations in the instrumental response.
For example, the \emph{Euclid} mission has implemented dedicated cleanliness and contamination-control programs to mitigate water ice and other molecular contaminants \citep{Euclid2023_IceContamination2023}, while the \emph{JWST} enforces extremely stringent molecular contamination limits for its optical surfaces.
These examples demonstrate that contamination by volatile species is a well-established cause of long-term throughput loss in space telescopes.

The VT CCD detectors operate at low temperatures to suppress dark current, making them particularly susceptible to optical contamination, unlike the VT main body and other optical components, which operate at room temperature.
To test the hypothesis of contamination, CCD bake-out operations were performed during the early commissioning phase.
During bake-out, the detector temperature was significantly elevated, which could partially mitigate contamination effects.

Following the first CCD bake-out on August 14, 2024, the photometric zero point showed a substantial recovery, corresponding to an approximate $40\%$ increase in throughput.
At the same time, noticeable changes were observed in the flat-field structures~\citep{Qiu2026}, indicating a redistribution of material within the optical system.
The strong and immediate photometric response to bake-out provides compelling evidence that the dominant contaminant is volatile in nature, most likely water ice.

In principle, a contamination layer could introduce wavelength-dependent absorption or scattering, leading not only to a gray throughput loss but also to color-dependent photometric effects. 
To assess this possibility, we performed a preliminary analysis using Gaia XP spectra of stars and examined the correlation between stellar color and zero-point variations after correcting for the long-term evolution. 
Within the current photometric precision, no significant color-dependent trend is found for either VT$\_B$ or VT$\_R$, indicating that the throughput degradation is consistent with a gray response change at the present accuracy level.
A more detailed spectrophotometric analysis, incorporating full time-dependent modeling, will be presented in a future dedicated calibration study.

Although a later bake-out (on November 15, 2005) continued to induce measurable responses in the zero point, its effectiveness diminished with time, producing only transient improvements that dissipated within a few days.
This behavior suggests that the contaminant was not fully removed but rather redistributed or re-condensed on other cold surfaces, eventually reaching a quasi-stable configuration.

Although the contamination remains a concern, our analysis shows it can be adequately controlled (Fig.~\ref{fig:stds_lc}).
Encouragingly, since early 2025, the VT zero points have reached a stable equilibrium, remaining consistent to within $\sim2\%$ over long timescales.
Nevertheless, continuous standard-star monitoring and regular calibration updates are still necessary to accurately track and correct any residual effects, ensuring that the scientific performance of VT is not significantly compromised.

\subsection{Filter System Performance: Red and Blue Leakage}
\label{subsec:leak}
The performance of a broadband photometric system can be affected by out-of-band spectral leakage, whereby light outside the nominal passband is transmitted through the filter and contributes to the measured flux.
For blue filters, so-called ``red leak'', contamination by long-wavelength photons, is of particular concern for intrinsically red sources.
Conversely, ``blue leak'' refers to short-wavelength contamination in red filters.
Such leakage introduces systematic photometric biases that depend on the SED of the observed source.

For the VT instrument, inspection of the synthetic transformations in Figure~\ref{fig:vt_trans_gaia},~\ref{fig:vt_trans_sdss} and~\ref{fig:vt_trans_bessell} suggests small but non-negligible responses outside the nominal wavelength ranges of both filters, 
implying the presence of out-of-band leakage.
Quantifying the red leak in VT$\_B$ and the blue leak in VT$\_R$ is therefore essential for assessing the robustness of the broadband photometry.

To evaluate the leakage properties, we adopted the same synthetic-photometry framework described in Section~\ref{subsec:zp}.
A grid of stellar spectra spanning a wide range of effective temperatures, from hot O-type to cool M-type main sequence stars, was used to compute synthetic fluxes based on the full VT response curves.
For each spectrum, we calculated the fractional leakage by:
\begin{equation}
    f_{\rm leak} = \frac{F_{\rm leak}}{F_{\rm total}} \times 100\%,
\end{equation}
where $F_{\rm total}$ is the integrated flux over the full response range ($3400-10000$\,\AA), and $F_{\rm leak}$ represents the flux transmitted outside the primary bandpass.
For VT$\_B$, red leak was defined as flux at $\lambda\ge7000$\,\AA, while for VT$\_R$ the blue leak was quantified as flux at $\lambda\le 6000$\,\AA.

\begin{figure}
\centering
\includegraphics[width=0.45\textwidth, angle=0]{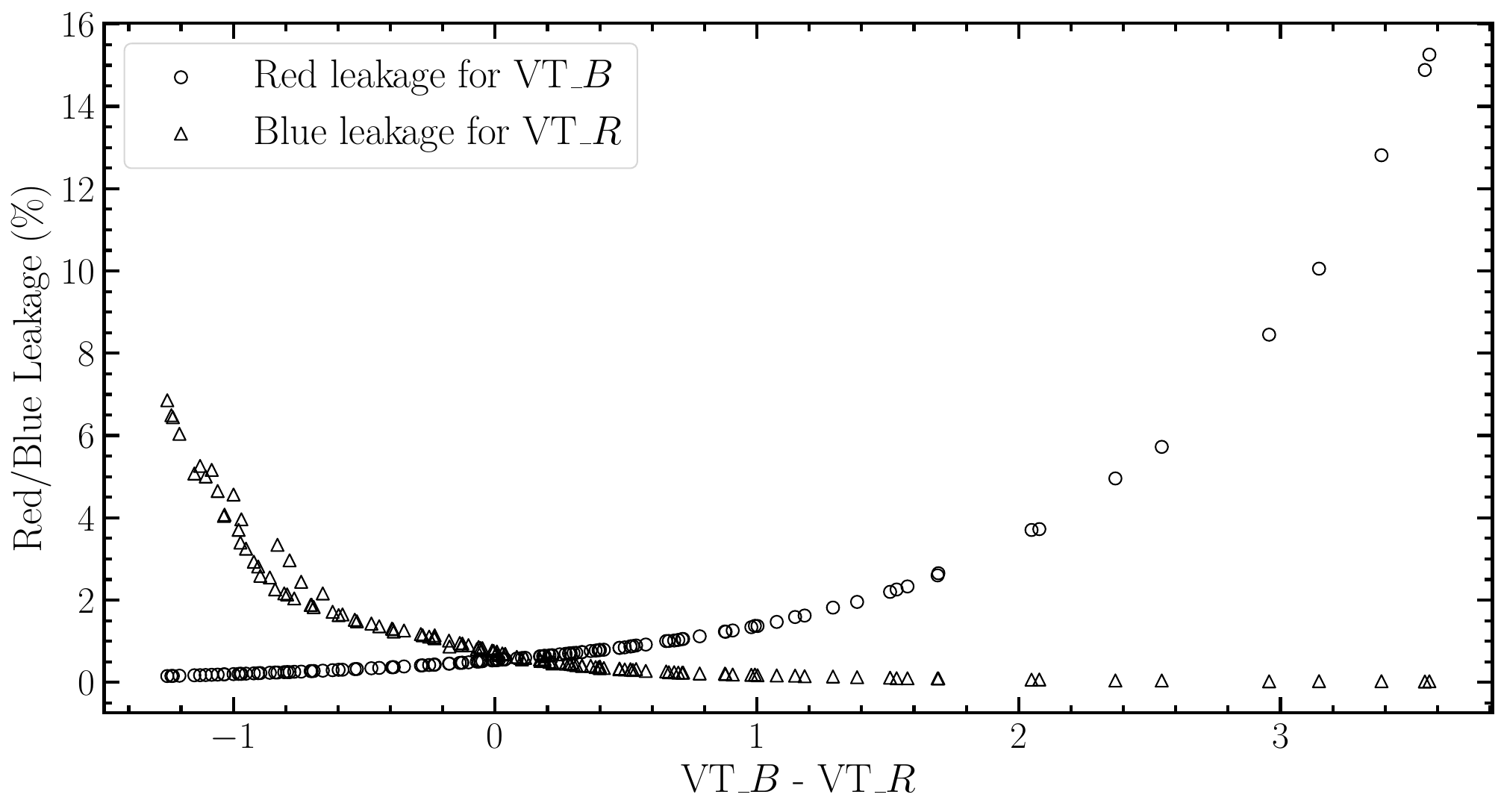}
\caption{Fractional red leakage in VT$\_B$ (circles) and blue leakage in VT$\_R$ (triangles) as a function of synthetic color $(\mathrm{VT}\_B - \mathrm{VT}\_R)$.}
\label{fig:leak_vt}
\end{figure}

Figure~\ref{fig:leak_vt} shows the resulting leakage fractions as a function of the synthetic color $(\mathrm{VT}\_B-\mathrm{VT}\_R)$, which serves as a proxy for stellar temperature.
The red-leak fraction of VT$\_B$ (blue circles) shows a strong dependence on color.
For hot, blue stars the leakage is negligible ($f_{\rm leak}\lesssim0.2\%$), whereas for increasingly red stars the fraction rises rapidly, reaching $\sim5\%$ at $(\mathrm{VT}\_B-\mathrm{VT}\_R)\approx2.5$ and exceeding $10-15\%$ for extremely cool stars.
This behavior indicates that the VT$\_B$ band remains effectively clean for most stellar types, but red leak becomes non-negligible for late-K and M-type stars.

In contrast, the blue-leak fraction of VT$\_R$ remains very small for the majority of spectral types.
For stars with $(\mathrm{VT}\_B-\mathrm{VT}\_R)\gtrsim 0$, the leakage fraction is below $1\%$ and typically at the $\sim0.1\%$ level, consistent with the designed red-band cutoff of VT$\_R$.
Toward the blue end of the sample, where $(\mathrm{VT}\_B-\mathrm{VT}\_R)\approx-1$, the blue leak increases modestly to $2-7\%$, reflecting enhanced transmission in the short-wavelength wing of the VT$\_R$ response.
Although this effect is much weaker than the red leak observed in VT$\_B$, it indicates that photometry of very hot stars may experience a small but measurable bias in VT$\_R$.

Overall, the leakage characteristics of the VT filters are well within acceptable limits for broadband photometry.
Red leak in VT$\_B$ becomes significant only for very cool stars, while blue leak in VT$\_R$ remains at the $\lesssim1\%$ level for most sources and exceeds a few percent only for extremely blue stars.
Consequently, leakage-induced photometric biases are expected to be smaller than the typical VT measurement uncertainties for the vast majority of astrophysical targets.

In a broader context, comparisons with other space-based photometric systems indicate that the VT filters perform competitively.
Several instruments, such as \emph{HST}/WFC3 \citep{Brown2008} and \emph{Swift}/UVOT \citep{Poole2008_SwiftUVOTphotcali,Breeveld2011_UVOTcal}, are known to exhibit noticeable red leak in their UV and blue filters, with contamination levels ranging from a few tenths of a percent up to $5-10\%$ for very red sources (e.g., WFC3/UVIS F225W and \emph{Swift} UVW2).
Compared to these systems, the VT$\_B$ red leak is of similar magnitude to the UV/blue filter leaks on \emph{HST} and \emph{Swift} when applied to extremely cool stars, yet remains negligible for hotter stars.
The VT$\_R$ blue leak, in contrast, is substantially smaller than that observed in many UV-optimized space-based filters, remaining below $1\%$ for most sources.
These comparisons reinforce the conclusion that the VT filter system achieves excellent overall performance, with leakage effects expected to be minor for typical scientific applications.

\section{Conclusion}
\label{sec:sum}
We present a comprehensive analysis of \textit{SVOM}/VT's in-orbit calibration, characterizing its astrometric performance, photometric system stability, and transformation relations with other photometric systems. The main results of this work can be summarized as follows:
\begin{itemize}
\item \textit{Astrometric performance:}
Using \textit{Gaia}~DR3 as the astrometric reference, VT achieves positional residuals better than 0.03$''$ for bright sources.
For faint sources, the expected astrometric accuracy of $\sim0.25''$ remains sufficient for reliable localization of GRB optical counterparts and transient sources.

\item \textit{Photometric response and zero points:}
The effective wavelengths, bandwidths, and AB zero points of the VT$\_B$ and VT$\_R$ bands were determined based on ground-based measurements and synthetic photometry.
The derived pre-launch AB zero points are 23.37~mag for VT$\_B$ and 23.22~mag for VT$\_R$.

\item \textit{Photometric stability:}
Continuous monitoring of spectrophotometric standard stars demonstrates that the VT photometric system remains stable at the 0.02~mag level since 2025 in orbit.
Short-term sensitivity variations associated with CCD bake-out operations are detected but are transient, with the zero points returning to their pre-bake-out levels within a few days.

\item \textit{Transformations with external photometric systems:}
Based on synthetic photometry using a broad stellar spectral library, we derived color-dependent transformations between the VT system and the \textit{Gaia}~DR3, SDSS, and Johnson--Cousins photometric systems.
Within the calibrated color ranges, the typical uncertainties of these transformations are $\sim0.03$~mag, enabling reliable cross-comparisons between VT data and existing photometric catalogs.

\item \textit{Out-of-band leakages:}
An assessment of out-of-band leakage shows that VT$\_B$ exhibits measurable red leak for very cool stars, while VT$\_R$ shows modest blue leak only for the bluest spectral types.
For the vast majority of astrophysical sources, the leakage-induced photometric biases are below the typical VT measurement uncertainties and therefore negligible for routine broadband applications.
\end{itemize}
Taken together, these results demonstrate that VT delivers stable and well-calibrated astrometric and photometric performance in orbit.
The established calibration framework ensures that VT data can be robustly integrated with multi-wavelength observations and fully supports the time-domain and transient science objectives of the \textit{SVOM} mission.

\begin{acknowledgements}
The Space-based multi-band astronomical Variable Objects Monitor (\textit{SVOM}) is a joint Chinese-French mission led by the Chinese National Space Administration (CNSA), the French Space Agency (CNES), and the Chinese Academy of Sciences (CAS).
We gratefully acknowledge the unwavering support of NSSC, IAMCAS, XIOPM, NAOC, IHEP, CNES, CEA, and CNRS.
This work is supported by the National Key R\&D Program of China (grant Nos. 2024YFA161170* and 2024YFA1611700) and by the National Natural Science Foundation of China (grant Nos. 12494571 and 12494570, 12494573, 12133003).
The authors are thankful for support from the Strategic Priority Research Program of the Chinese Academy of Sciences (Grant No.XDB0550401).
\end{acknowledgements}

\appendix                  

\bibliographystyle{raa}
\bibliography{ref}




\label{lastpage}

\end{document}